\documentclass[showpacs,preprintnumbers,eqsecnum,prd,nofootinbib,twocolumn]{revtex4}  

\usepackage{graphicx}

\def\beq{\begin{equation}}
\def\eeq{\end{equation}}
\def\rmd{{\rm d}}

\begin{document}

\title{Effect of an arbitrary spin orientation on the quadrupolar structure of an extended body in a Schwarzschild spacetime}

\author{Donato Bini and Andrea Geralico}
  \affiliation{
Istituto per le Applicazioni del Calcolo ``M. Picone,'' CNR, I-00185 Rome, Italy
}

\date{\today}

\begin{abstract}
The influence of an arbitrary spin orientation on the quadrupolar structure of an extended body moving in a Schwarzschild spacetime is investigated.
The body dynamics is described by the Mathisson-Papapetrou-Dixon model, without any restriction on the motion or simplifying assumption on the associated spin vector and quadrupole tensor, generalizing previous works.
The equations of motion are solved analytically in the limit of small values of the characteristic length scales associated with the spin and quadrupole variables
with respect to the characteristic length of the background curvature. 
The solution provides all corrections to the circular geodesic on the equatorial plane taken as the reference trajectory due to both dipolar and quadrupolar structure of the body as well as the conditions which the nonvanishing components of the quadrupole tensor must fulfill in order that the problem be self-consistent.
\end{abstract}

\pacs{04.20.Cv}

\maketitle

\section{Introduction}

An extended test body moving in a given gravitational background is commonly described according to the Mathisson-Papapetrou-Dixon (MPD) model \cite{math37,papa51,tulc59,dixon64,dixon69,dixon70,dixon73,dixon74,dixon_varenna}.
The MPD equations of motion involve a reference world line, properly set up to represent the center-of-mass line of the body, and a number of vector and tensor fields defined along it through a multipole moment expansion, similarly to the standard nonrelativistic theory. 
The model is fully determined and self-consistent at the dipolar order, i.e., for an extended body endowed with spin only, providing a set of evolution equations for both the linear and angular momentum of the body.
For the quadrupolar as well as higher multipolar orders, instead, there are no evolution equations for the quadrupole and higher multipole tensors, and their evolution is fixed entirely by the body's internal dynamics.
The contribution of higher multipoles appears in the form of additional force and torque terms.
Therefore, one has to supply the structure of the body as external information, e.g., specifying the equation of state of its matter-energy content. 
This fact represents a peculiarity of the model itself, which allows for many different approaches. 

According to Dixon's construction, the quadrupole tensor shares the same symmetries of the Riemann tensor and is completely specified by two symmetric and trace-free spatial tensors, i.e., the mass quadrupole (electric-type) and the current quadrupole (magnetic-type) tensors.
The most natural and simplifying choice consists in considering the body as \lq\lq quasi-rigid,'' i.e., all unspecified quantities describing its shape are taken constant in the body-fixed frame, i.e., adapted to the 4-momentum of the body itself \cite{ehlers77}. 
Alternatively, one can assume the quadrupole tensor be directly related to the Riemann tensor, having the same symmetry properties.
For instance, in Ref. \cite{steinhoff} the electric and magnetic parts of the quadrupole tensor have been taken proportional to the electric and magnetic parts of the Riemann tensor, respectively, to study quadrupole deformation effects induced by the tidal field of a black hole on the motion of a spinning body.
One can also require that the structure of the body be completely determined by its spin, with a quadratic-in-spin quadrupole tensor, which is of purely electric type, being proportional to the trace-free part of the square of the spin tensor \cite{steinhoff2,hinderer,bfg}.
The interest in such an approach is the possibility to include spin-induced quadrupole corrections in the post-Newtonian dynamics of a two-body system (see, e.g., Refs. \cite{porto06,porto08,steinhoff08,hergt08,steinhoff09,Levi:2014gsa,Levi:2015msa}).

Beside the choice of the quadrupole tensor, further simplifications to the problem come from the symmetries of the considered gravitational field as well as the particular kind of motion one is interested in.  
In the literature, the motion of extended bodies around a compact object is usually assumed to be confined to the equatorial plane of reflection symmetric spacetimes (like Schwarzschild and Kerr backgrounds) taken as the orbital plane, the spin vector being constant in magnitude and necessarily directed orthogonally to it. This is a highly symmetric situation which imposes a strong restriction on the structure of the body, reducing the number of nonvanishing components of the quadrupole tensor.
The latter has in general 20 independent components, but only 10 of them are actually relevant, namely those obeying the symmetries of the MPD equations, as shown in Ref. \cite{quadrup_kerr1}.
In fact, the quadrupole tensor enters the MPD equations only through certain contractions with the Riemann tensor and its covariant derivative.
This is also expected from the standard post-Newtonian formulation of motion of many-body systems.
When the motion is restricted to the equatorial plane, the number of nonvanishing effective components of the quadrupole tensor then reduces from 10 to 5, 3 belonging to the mass quadrupole moment, and 2 to the current quadrupole moment \cite{quadrup_kerr1}.

In the present paper we will study the dynamics of an extended body in a Schwarzschild spacetime in the framework of the MPD model, without any simplifying assumption on the body dynamics and structure, allowing it to move off the equatorial plane with an arbitrary orientation of the spin vector, generalizing previous results \cite{quadrup_schw}. 
Following the MPD prescriptions, we only require that the body does not perturb significantly the background field, so that backreaction effects can be neglected and the body structure only produces very small deviations from geodesic motion.
This condition is indeed implicit in the MPD model, and allows to treat the equations of motion perturbatively, in the sense that the natural length scales associated with the body, i.e., the \lq\lq bare'' mass as well as the spin and quadrupole characteristic lengths, are taken to be small enough if compared with the length scale associated with the background curvature.
The resulting simplified set of differential equations can be integrated analytically.
Initial conditions are chosen so that the world line of the extended body has the same starting point and is initially tangent to a timelike circular geodesic on the equatorial plane, taken as a reference world line.
It is known that the presence of a spin component in the equatorial plane induces an oscillation of the body path in and out of the equatorial plane to first order in spin (i.e., taking into account only corrections which are linear in spin) \cite{mashsingh,spin_dev_schw}. The orthogonal component is instead responsible for an oscillating behavior of the radial distance about the reference radius, while the azimuthal motion undergoes similar oscillations plus an additional secular drift.
We will show that second order corrections to an initially circular geodesic motion (i.e., spin squared and mass quadrupole terms) introduce secular effects in both radial and polar motions enhancing deviations.

Furthermore, we will discuss the consequences of that general situation on the quadrupolar structure of the body. 
The spin evolution equations, indeed, provide some compatibility conditions involving spin vector components, first order corrections to the orbit and components of the quadrupole tensor, which turn out to be varying with time.  
As a result, the mass quadrupole moment associated with the body changes during the evolution, causing its shape to change too, passing from nearly spherical to highly deformed configurations.
A variable mass quadrupole moment is usually generated in a binary system because of the tides produced by the higher mass. In such a situation the net gravitational radiation associated with the motion of the smaller mass is due to its orbit, the time varying tides and the interference between them \cite{mash1,mash2}. 
Furthermore, changes in the gravitational quadrupole moment of the companion star are expected to account for most of the observed variations in the orbital parameters of binary pulsar systems \cite{damour,kopeikin}. Noteworthy, recent radio timing observations of the eclipsing millisecond binary pulsar PSR J2051-0827 have provided evidences for variations of the quadrupole moment in its companion \cite{applegate,doroshenko,lazaridis}. Such variations together with the spin precession of the companion star have been shown to be responsible for the changes of the orbital period, inclination angle and projected semimajor axis of the binary system.  
Although the underlying mechanism causing a varying quadrupole moment is most likely of non-gravitational nature (e.g., driven by the magnetic activity in close binaries \cite{applegate,lanza}), the purely gravitational effect discussed here may play a role.

We will follow notations and conventions of Ref. \cite{MTW}. 
Units are chosen so that $G=1=c$ and the metric signature is $-+++$.
Greek indices run from $0$ to $3$, whereas Latin indices from $1$ to $3$.

\section{MPD equations in the quadrupole approximation}

Consider an extended body endowed with structure up to the quadrupole moment. 
The MPD equations are
\begin{eqnarray}
\label{papcoreqs1}
\frac{{\rm D}P^{\mu}}{\rmd \tau} & = &
- \frac12 \, R^\mu{}_{\nu \alpha \beta} \, U^\nu \, S^{\alpha \beta}
-\frac16 \, \, J^{\alpha \beta \gamma \delta} \, \nabla^\mu R_{\alpha \beta \gamma \delta}
\nonumber\\
& \equiv & F^\mu_{\rm (spin)} + F^\mu_{\rm (quad)} \,,
\\
\label{papcoreqs2}
\frac{{\rm D}S^{\mu\nu}}{\rmd \tau} & = & 
2 \, P^{[\mu}U^{\nu]}+
\frac43 \, J^{\alpha \beta \gamma [\mu}R^{\nu]}{}_{\gamma \alpha \beta}
\nonumber\\
&\equiv & D^{\mu \nu}_{\rm (spin)} + D^{\mu \nu}_{\rm (quad)} \,,
\end{eqnarray}
where $P^{\mu}=m u^\mu$ (with $u \cdot u = -1$) is the total 4-momentum of the body with mass $m$, $S^{\mu \nu}$ is a (antisymmetric) spin tensor, $J^{\alpha\beta\gamma\delta}$ is the quadrupole tensor, and $U^\mu=\rmd z^\mu/\rmd\tau$ is the timelike unit tangent vector of the \lq\lq center of mass line'' (with parametric equations $x^\mu=z^\mu(\tau)$) used to make the multipole reduction, parametrized by the proper time $\tau$. 

In order the model to be mathematically self-consistent the following additional conditions should be imposed \cite{tulc59,dixon64}
\beq
\label{tulczconds}
S^{\mu\nu}u{}_\nu=0\,.
\eeq
Consequently, the spin tensor can be fully represented by a spatial vector (with respect to $u$),
\beq
S(u)^\alpha=\frac12 \eta(u)^\alpha{}_{\beta\gamma}S^{\beta\gamma}
\,,
\eeq
where $\eta(u)_{\alpha\beta\gamma}=\eta_{\mu\alpha\beta\gamma}u^\mu$ is the spatial (with respect to $u$) unit volume 3-form with $\eta_{\alpha\beta\gamma\delta}=\sqrt{-g} \epsilon_{\alpha\beta\gamma\delta}$ the unit volume 4-form and $\epsilon_{\alpha\beta\gamma\delta}$ ($\epsilon_{0123}=1$) the Levi-Civita alternating symbol. 
It is also useful to introduce the signed magnitude $s$ of the spin vector
\beq
\label{sinv}
s^2=S(u)^\beta S(u)_\beta = \frac12 S_{\mu\nu}S^{\mu\nu}\,, 
\eeq
which is in general not constant along the trajectory of the extended body. 

The quadrupole tensor $J^{\alpha\beta\gamma\delta}$ by its definition has the same algebraic symmetries as the Riemann tensor, but enters the MPD equations only through certain combinations, which reduce the number of effective components from 20 to 10 \cite{quadrup_schw,quadrup_kerr1,quadrup_kerr_num}.
Therefore, in complete analogy with the $1+3$ splitting of the Riemann tensor with respect to a given timelike congruence, it can be written in the form 
\begin{eqnarray}
\label{deco_bar_u5}
J^{\alpha\beta}{}_{\gamma\delta}&=&4 u^{[\alpha}[X(u)]^{\rm STF}{}^{\beta]}{}_{[\gamma} u_{\delta]}\nonumber\\
&&
+2  u^{[\alpha}[W(u)]^{\rm STF}{}^{\beta]}{}_\sigma \eta(u)^{\sigma}{}_{ \gamma\delta}\nonumber\\
&&
+2 u_{[\gamma}[W(u)]^{\rm STF}{}_{\delta]}{}_\sigma \eta(u)^{\sigma \alpha\beta}\,,
\end{eqnarray} 
where $X(u)$ and $W(u)$ are symmetric and trace-free (STF) spatial tensors as \lq\lq measured'' by an observer comoving with the body, representing the mass quadrupole moment and the flow (or current) quadrupole moment, respectively (see, e.g., Ref. \cite{ehlers77}).

In stationary and axisymmetric spacetimes endowed with Killing symmetries 
the total energy $E$ and the angular momentum $J$ are conserved quantities
along the motion associated with the timelike Killing vector $\xi=\partial_t$
and the azimuthal Killing vector $\eta=\partial_\phi$, respectively. 
They are given by
\begin{eqnarray}
\label{totalenergy}
E&=&-\xi_\alpha P^\alpha +\frac12 S^{\alpha\beta}F^{(t)}_{\alpha\beta}\,,\nonumber\\
J&=&\eta_\alpha P^\alpha -\frac12 S^{\alpha\beta}F^{(\phi)}_{\alpha\beta}\,,
\end{eqnarray}
where
\beq
F^{(t)}_{\alpha\beta}=\nabla_\beta \xi_\alpha=g_{t[\alpha,\beta]}\,, \quad
F^{(\phi)}_{\alpha\beta}=\nabla_\beta \eta_\alpha=g_{\phi[\alpha,\beta]}\,, 
\eeq
are the Papapetrou fields associated with the Killing vectors. Note that $E$
and $J$ as defined above are conserved quantities to all multipole orders in
spite of the higher multipolar structure of the body \cite{ehlers77}.

\subsection{Perturbative approach}

Consider a pair of world lines emanating from a common spacetime point, one a geodesic with 4-velocity $U_{(\rm geo)}$, the other the world line of an extended body deviating from the reference one because of the combined effects of both the spin-curvature and quadrupole-curvature couplings, with 4-velocity $U$. 
Introduce a smallness indicator $\epsilon\ll1$ to distinguish between the order of multipolar approximation, so that $S^{\mu\nu}=O(\epsilon)$ and $J^{\alpha\beta\gamma\delta}=O(\epsilon^2)$.
Solutions to the MPD equations can then be found in the general form
\begin{eqnarray}
\label{U_pert_v}
x^\alpha&=&x^\alpha_{\rm (geo)}+\epsilon x^\alpha_{(1)}+\epsilon^2x^\alpha_{(2)}\,, \nonumber\\
U&=&U_{\rm (geo)}+\epsilon U_{(1)}+\epsilon^2U_{(2)}\,.
\end{eqnarray}
The mass $m=(-P^\mu P_\mu)^{1/2}$ of the body is a conserved quantity to first order and the 4-momentum vector $P$ is parallel to the 4-velocity $U$, so that one can assume
\begin{eqnarray}
\label{u_pert_v}
m&=&m_0+\epsilon^2m_{(2)}\,, \nonumber\\
u&=&U_{\rm (geo)}+\epsilon U_{(1)}+\epsilon^2u_{(2)}\,,
\end{eqnarray}
where $m_0$ denotes the \lq\lq bare'' mass.
The second order correction to the mass of the body turns out to be
\beq
\label{m2_def}
m_{(2)}=\frac16 J^{\alpha \beta \gamma \delta} R_{\alpha \beta \gamma \delta}\,,
\eeq
whereas the unit vectors $U$ and $u$ are related by
\beq
\label{reluUgen}
u^\mu=U^\mu+\frac1{m_0}D^{\mu \nu}_{\rm (quad)}U_\nu
+\frac1{m_0^2}S^{\mu\nu} F_{\rm (spin)}{}_{\nu}+O(\epsilon^3)\,,
\eeq 
providing four algebraic relations between their components.
Substituting the expansions above into the MPD equations (\ref{papcoreqs1}) and (\ref{papcoreqs2}) then leads to two different sets of evolution equations for the first order and second order quantities, respectively, neglecting terms of higher order. 

It is worth noting that $U_{\rm (geo)}$ and $U$ are unit tangent vectors to different timelike world lines, which are parametrized by different proper times: hence, one should use $\tau_{\rm (geo)}$ as the proper time parameter along $U_{\rm (geo)}$ and $\tau$ as the proper time parameter along $U$. 
However, from their definitions,
\beq
d\tau=-U_\alpha dx^\alpha\,,\qquad 
d\tau_{\rm (geo)}=-U_{\rm (geo)}{}_\alpha dx_{\rm (geo)}^\alpha
\eeq
and recalling the normalization condition $U\cdot U=-1$, one obtains that $\tau$ and $\tau_{\rm (geo)}$ can be identified to the second order of approximation, i.e.,
\beq
\tau=\tau_{\rm (geo)}+O(\epsilon^3)\,.
\eeq
Therefore, although the two world lines are parametrized by different proper times, the latter are synchronized so that $\tau$ can be used unambiguously for that single proper time parametrization of both world lines.

We are interested here in solutions to the MPD equations which describe deviations from geodesic motion due to both the spin-curvature force and the quadrupolar force. 
Hence, we will choose initial conditions so that the world line of the extended body has the same starting point as the reference geodesic, i.e., 
\beq
\label{inicond1}
x^{\alpha}_{(1)}(0)=0=x^{\alpha}_{(2)}(0)\,.
\eeq
The two world lines in general have not a common unit tangent vector at $\tau_{\rm (geo)}=0=\tau$; as $\tau$ increases, then, they deviate from each other.
We will require that the 4-velocity $U$ be initially tangent to the geodesic 4-velocity $U_{\rm (geo)}$, which implies in addition 
\beq
\label{inicond2}
\frac{\rmd x^{\alpha}_{(1)}(0)}{\rmd\tau}=0=\frac{\rmd x^{\alpha}_{(2)}(0)}{\rmd\tau}\,.
\eeq
In the next section we will specialize our analysis to the Schwarzschild background.

\section{Dynamics of extended bodies in a Schwarzschild spacetime}

Consider the Schwarzschild spacetime in standard coordinates $(t,r,\theta,\phi)$, with line element written in standard form as
\beq
\label{sch}
\rmd s^2=-N^2\rmd t^2+N^{-2}\rmd r^2+r^2(\rmd\theta^2+\sin^2\theta \rmd\phi^2)\,,
\eeq
where $N$ denotes the lapse function 
\beq
N=\sqrt{1-\frac{2M}{r}}\,.
\eeq
An orthonormal frame adapted to the static observers, i.e., those following the coordinate time lines with $4$-velocity $n=N^{-1}\partial_t$, is given by 
\begin{eqnarray}
\label{frame}
e_{\hat t}&=&n\,, \qquad
e_{\hat r}=N\partial_r\,,\nonumber\\
e_{\hat \theta}&=&\frac{1}{r}\partial_\theta\,,\qquad 
e_{\hat\phi}=\frac{1}{r\sin\theta}\partial_\phi\,,
\end{eqnarray}
with dual $\omega^{\hat t}=N\rmd t$, $\omega^{\hat r}=N^{-1}\rmd r$, $\omega^{\hat \theta}=r\rmd \theta$ and $\omega^{\hat \phi}=r\sin\theta\rmd \phi$.

Let the reference world line be a circular geodesic in the equatorial plane at radius $r = r_0$.
The associated $4$-velocity $U_{\rm (geo)}=U_K$ is 
\beq
\label{Ugeocirc}
U_K=\gamma_K (n\pm \nu_K e_{\hat \phi})=\Gamma_K (\partial_t \pm \zeta_K \partial_\phi)\,,
\eeq
where the $\pm$ signs refer to co-rotating $(+)$ and counter-rotating $(-)$ motion with respect to increasing values of the azimuthal coordinate, respectively.
Here $\zeta_K>0$ and $\nu_K>0$ (with associated Lorentz factor $\gamma_K=(1-\nu^2_K)^{-1/2}$) denote the Keplerian angular velocity and linear velocity, respectively, and $\Gamma_K$ is a normalization factor defined by
\begin{eqnarray}
\label{kepler}
\zeta_K&=&\sqrt{\frac{M}{r_0^3}}\,, \qquad
\nu_K=\sqrt{\frac{M}{r_0-2M}}\,,\nonumber\\
\gamma_K&=&\sqrt{\frac{r_0-2M}{r_0-3M}}\,, \qquad
\Gamma_K=\frac{\gamma_K}{N}=\frac{\gamma_K\nu_K}{r_0\zeta_K}
\,.
\end{eqnarray}
The circular geodesic is thus described by the parametric equations
\begin{eqnarray}
t_{\rm (geo)} &=&  t_0+\Gamma_K \tau \,,\qquad
r_{\rm (geo)} = r_0\,,\nonumber\\
\theta_{\rm (geo)} &=&\frac{\pi}{2}\,,\qquad
\phi_{\rm (geo)} =  \phi_0\pm\Gamma_K\zeta_K \tau \,.
\end{eqnarray}
It is useful to introduce the unit vector $\bar U_K$ along the azimuthal direction in the local rest space of the circular geodesic, orthogonal to $E_0\equiv U_K$ in the $t$-$\phi$ plane, i.e.,
\beq
\bar U_K=\gamma_K ( \pm e_{\hat \phi}+\nu_K e_{\hat t})\,.
\eeq
An orthonormal frame adapted to $U_K$ is thus given by
\beq
\label{geoframe}
E_1=e_{\hat r}\,,\qquad
E_2=\bar U_K\,,\qquad
E_3=-e_{\hat \theta}\,,
\eeq
with $E_3$ aligned with the (positive)  $z$-axis of a naturally defined Cartesian frame. 

Finally, the circular geodesic conserved energy and angular momentum are given by
\beq
\label{geoEJ}
E_K=m_0N\gamma_K\,,\qquad
J_K=\pm m_0r_0\gamma_K\nu_K\,,
\eeq
respectively.

\subsection{First order solution}

To first order the set of MPD equations (\ref{papcoreqs1}) and (\ref{papcoreqs2}) reduces to
\begin{eqnarray}
\label{pap_small_spin}
m\frac{DU_{(1)}^{\mu}}{\rmd \tau} &=& F^{\rm (spin)}{}^\mu+O(\epsilon^2)\,, \nonumber\\
\frac{DS^{\mu\nu}}{\rmd \tau} &=& O(\epsilon^2)\,.
\end{eqnarray}
The spin vector must be orthogonal to $U$ due to the supplementary conditions (\ref{tulczconds}), so that
\beq
S = S^{\hat r}e_{\hat r} 
          +S^{\hat \theta}e_{\hat \theta}
          \pm \gamma_K^{-1} S^{\hat \phi}\bar U_K\,,
\eeq
and turns out to be parallel transported along the reference circular geodesic due to the spin evolution equations.
That leads to a simple rotation of the spin components in the $r$-$\phi$ plane within the local rest space of the circular geodesics.
The corresponding solution can then be written as 
\beq
\label{final_spin}
S=s_\Vert [\cos \alpha e_{\hat r}\pm \sin \alpha \bar U_K]-s_\perp e_{\hat \theta}
\,,
\eeq
where a polar representation for the spin vector has been conveniently introduced such that $S^{\hat r}(\tau)=s_\Vert\cos\alpha$, $S^{\hat \phi}(\tau)=\gamma_Ks_\Vert\sin\alpha$ and $S^{\hat \theta}=-s_{\perp}$, with 
\beq
\label{sol_alpha}
\alpha(\tau)=\alpha_0\mp \zeta_K \tau \,.
\eeq
The quantities $s_{\Vert}=[S^{\hat r}(0)^2+S^{\hat \phi}(0)^2/\gamma_K^2]^{1/2}$ and $s_\perp$ are constant due to the conservation of the spin magnitude. 

The solution for the orbit is then given by \cite{spin_dev_schw}
\begin{eqnarray}
t_{(1)}
&=&\pm \frac{\nu_K^2}{\zeta_K}\phi_{(1)}
\,, \nonumber\\
r_{(1)}
&=& \pm r_0\Sigma_\perp(1-\cos\Omega_{\rm(ep)}\tau)
\,, \nonumber\\ 
\theta_{(1)}
&=&  
\mp\Sigma_\parallel\left[
\cos\alpha-\cos\alpha_0\cos\Omega_{\rm(orb)}\tau\right.\nonumber\\
&&\left.
\mp\frac1{\Gamma_K}\sin\alpha_0\sin\Omega_{\rm(orb)}\tau
\right]\,,
\nonumber \\
\phi_{(1)}
&=&
2\frac{\Omega_{\rm(orb)}}{\Omega_{\rm(ep)}}
\Sigma_\perp(\sin\Omega_{\rm(ep)}\tau - \Omega_{\rm(ep)}\tau)
\,,
\end{eqnarray}
where
\beq
\Sigma_\perp = 3N^2(M\zeta_K)\frac{\Omega_{\rm(orb)}^2}{\Omega_{\rm(ep)}^2} \sigma_\perp\,, \qquad
\Sigma_\Vert = N(r_0\zeta_K)\sigma_{\Vert}\,,
\eeq
and  the dimensionless spin quantities
\beq
\sigma_\Vert=\frac{s_\Vert}{m_0M}  \,,\qquad 
\sigma_\perp=\frac{s_\perp}{m_0M}\,
\eeq
have been introduced.
Here
\begin{eqnarray}
\label{frequencies}
  \Omega_{\rm(ep)} &\equiv& \sqrt{\frac{M (r_0-6M)}{r_0^3 (r_0 -3M)}} \,,\nonumber\\
  \Omega_{\rm(orb)} &\equiv&\Gamma_K\zeta_K
	=\frac{1}{r_0}\sqrt{\frac{M}{r_0-3M}}
\end{eqnarray}
are respectively the well known epicyclic frequency governing the radial perturbations of circular geodesics and the orbital frequency governing the geodesic oscillations out of the equatorial plane. The latter frequency together with the spin-precession frequency due to the spin oscillation driving term governs the polar angle oscillations about the equatorial plane.
Their ratio 
\beq
\frac{ \Omega_{\rm(orb)}}{ \Omega_{\rm(ep)}}=\left(1-\frac{6M}{r_0}\right)^{-1/2}
\eeq
will enter most of the relations below, also implying the allowed range for radial distance $r_0\ge 6M$.

The solution for $U$ (which at $\tau=0$ is aligned with the circular geodesic at $r_0$) is then given by $U=U_K+U_{(1)}$, with
\begin{eqnarray}
\label{Usol}
U_{(1)}&=&\pm  \nu_K \Sigma_\perp \left[\frac{\Omega_{\rm(ep)}}{\zeta_K}\sin\Omega_{\rm(ep)}\tau\,e_{\hat r}\right.\nonumber\\
&&\left.
+2\left(\cos\Omega_{\rm(ep)}\tau-1\right)\bar U_{\rm (geo)}\right]\nonumber\\
&&+ (r_0\zeta_K) \Sigma_\Vert \left[\sin \alpha_0 \cos\Omega_{\rm(orb)}\tau\right.\nonumber\\
&&\left.
 \mp  \Gamma_K\cos \alpha_0  \sin\Omega_{\rm(orb)}\tau
 -\sin\alpha\right]e_{\hat \theta}\,. \nonumber\\
\end{eqnarray} 
When decomposed with respect to the frame (\ref{frame}) adapted to the static observers, the latter writes as $U_{(1)}=U_{(1)}^{\hat \alpha}e_{\hat \alpha}$, leading to the following relations between frame and coordinate components 
\begin{eqnarray}
U_{(1)}^{\hat t}&=&NU_{(1)}^t+\Omega_{\rm(orb)}\nu_K r_{(1)}
=\pm\nu_KU_{(1)}^{\hat \phi}\,, \nonumber\\
U_{(1)}^{\hat r}&=&N^{-1}U_{(1)}^r\,, \nonumber\\
U_{(1)}^{\hat \theta}&=&r_0U_{(1)}^\theta\,, \nonumber\\
U_{(1)}^{\hat \phi}&=&r_0U_{(1)}^\phi\pm\Omega_{\rm(orb)}r_{(1)}\,.
\end{eqnarray} 

Finally, the first order corrections to the circular geodesic conserved energy and angular momentum (\ref{geoEJ}) are given by
\begin{eqnarray}
\label{costmoto1}
E_{(1)}&=&\pm m_0(r_0\zeta_K)^5\Gamma_K\sigma_\perp\,, \nonumber\\
J_{(1)}&=&m_0r_0(r_0\zeta_K)^2N^2\Gamma_K\sigma_\perp\,,
\end{eqnarray} 
respectively, as from Eq. (\ref{totalenergy}).

\subsection{Second order solution}

The evolution of the spin vector is completely determined by the first order equations. Therefore, the spin evolution equations (\ref{papcoreqs2}) to second order simply provide three algebraic relations between the components of $u_{(2)}$ and $U_{(2)}$ (in agreement with Eq. (\ref{reluUgen})) plus three compatibility conditions involving spin vector components, first order corrections to the orbit $x^\alpha_{(1)}$ and components of the quadrupole tensor.
In fact, 
\beq
D^{\mu \nu}_{\rm (spin)}=2\epsilon^2m_0(u_{(2)}-U_{(2)})^{[\mu}U_K^{\nu]}+O(\epsilon^3)\,,
\eeq
and $D^{\mu \nu}_{\rm (quad)}=O(\epsilon^2)$.
A further condition comes from the evolution equation for the second order correction (\ref{m2_def}) to the mass of the body.
Contracting Eq. (\ref{papcoreqs1}) with $u_\mu$ leads to
\beq
\frac{\rmd m_{(2)}}{\rmd \tau}=-U_{K\,\mu}F^\mu_{\rm (quad)}+O(\epsilon^3)\,,
\eeq
whereas contracting with $U_\mu$ yields
\beq
\frac{\rmd m_{(2)}}{\rmd \tau}=-U_{K\,\mu}F^\mu_{\rm (quad)}
+\frac16 R_{\alpha \beta \gamma \delta}\frac{{\rm D}J^{\alpha \beta \gamma \delta}}{\rmd \tau}+O(\epsilon^3)\,,
\eeq
being $F_{\rm (spin)}$ orthogonal to $U$, implying that 
\beq
\label{m2cond}
R_{\alpha \beta \gamma \delta}\frac{{\rm D}J^{\alpha \beta \gamma \delta}}{\rmd \tau}=O(\epsilon^3)\,,
\eeq
which involves certain components of the quadrupole tensor and their first derivatives with respect to proper time.

It is convenient to introduce the dimensionless frame components $\tilde S^{\hat a}\equiv S^{\hat a}/(m_0M)$ of the spin vector as well as the following combinations $K_i=K_i(\tau)$ of the quadrupole tensor components
\begin{eqnarray}
\label{Kdefs} 
K_1&=&2\tilde X_{11}+\tilde X_{22}\pm2\nu_K\frac{1-(r_0\zeta_K)^2}{(r_0\zeta_K)^2}\tilde W_{13}
\,,\nonumber\\
K_2&=&\tilde X_{12}\pm2\nu_K\tilde W_{23}
\,,\nonumber\\
K_3&=&2\tilde W_{12}\mp\nu_K\tilde X_{23}
\,,\nonumber\\
K_4&=&\pm4N(r_0\zeta_K)\tilde W_{13}+\left[1-(r_0\zeta_K)^2\right]\tilde X_{11}\nonumber\\
&&
+(r_0\zeta_K)^2\tilde X_{22}
\,,\nonumber\\
K_5&=&2(2\tilde W_{11}+\tilde W_{22})\mp\nu_K\frac{1-(r_0\zeta_K)^2}{(r_0\zeta_K)^2}\tilde X_{13}
\,,
\end{eqnarray} 
where
\beq
\tilde X_{ab}\equiv\frac{X(u)_{ab}}{m_0M^2}\,,\qquad
\tilde W_{ab}\equiv\frac{W(u)_{ab}}{m_0M^2}
\eeq
are dimensionless quadrupole quantities obtained by suitably rescaling the frame components of the electric and magnetic parts of the quadrupole tensor with respect to the frame (\ref{geoframe}) adapted to the circular geodesics.

The spin evolution equations (\ref{papcoreqs2}) then give
\begin{eqnarray}
(u_{(2)}-U_{(2)})^{\hat t}&=&(r_0\zeta_K)^9\frac{\Gamma_K^3}{\nu_K}\left\{
3\left[(\tilde S^{\hat r})^2-\sigma_\perp^2\right]-4K_1
\right\}\nonumber\\
&=&\pm\nu_K(u_{(2)}-U_{(2)})^{\hat\phi}
\,,\nonumber\\
(u_{(2)}-U_{(2)})^{\hat r}&=&\mp (r_0\zeta_K)^7\Gamma_K\left[
3\tilde S^{\hat r}\tilde S^{\hat \phi}-4\gamma_K K_2
\right]
\,,\nonumber\\
(u_{(2)}-U_{(2)})^{\hat \theta}&=&\mp (r_0\zeta_K)^7\Gamma_K\left[
3\sigma_\perp\tilde S^{\hat \phi}\pm4\frac{\gamma_K}{\nu_K} K_3
\right]
\,,\nonumber\\
\end{eqnarray} 
together with the following compatibility conditions
\begin{eqnarray}
\label{compatib}
K_2&=&-\frac{1}{4(r_0\zeta_K)^4\gamma_K^2}\left[
N\tilde S^{\hat r}U_{(1)}^{\hat \theta}
\pm (r_0\Omega_{\rm(orb)})\tilde S^{\hat \phi}\theta_{(1)}
\right]
\,,\nonumber\\
K_3&=&
\mp\frac{1}{4(r_0\zeta_K)^5N^2\Gamma_K^4}\left[
(r_0\Omega_{\rm(orb)})^2\tilde S^{\hat r}U_{(1)}^{\hat r}
+\sigma_\perp U_{(1)}^{\hat \theta}\right.\nonumber\\
&&\left.
-\tilde S^{\hat \phi}U_{(1)}^{\hat \phi}
\pm\frac{\nu_K}{\gamma_K}(1-(r_0\zeta_K)^2\Gamma_K^4)\tilde S^{\hat \phi}\frac{r_{(1)}}{r_0}
\right]
\,,\nonumber\\
K_5&=&\frac{1}{4(r_0\zeta_K)^5N\Gamma_K^3}\left[
\Gamma_K\nu_K\sigma_\perp\theta_{(1)}\right.\nonumber\\
&&\left.
\pm\tilde S^{\hat r}\left(
U_{(1)}^{\hat \phi}-6(r_0\zeta_K)^6\Gamma_K^3\sigma_\perp
-\frac{\nu_K}{\gamma_K}\frac{r_{(1)}}{r_0}
\right)
\right]
\,.\nonumber\\
\end{eqnarray} 

The second order correction to the mass of the body $m_{(2)}$ is given by Eq. (\ref{m2_def}), which reads
\beq
m_{(2)}=2m_0(M\Omega_{\rm(orb)})^2K_4\,.
\eeq
The condition (\ref{m2cond}) implies
\beq
r_0\frac{\rmd K_4}{\rmd \tau}=\pm2N^2(r_0\zeta_K) K_2\,,
\eeq
whose integration yields
\beq
r_0[K_4(\tau)-K_4(0)]=\pm2N^2(r_0\zeta_K) \int_0^\tau K_2(\xi)\rmd\xi\,.
\eeq
Therefore, the solution for $m_{(2)}$ turns out to be

\begin{figure} 
\typeout{*** EPS figure 1}
\begin{center}
\includegraphics[scale=0.4]{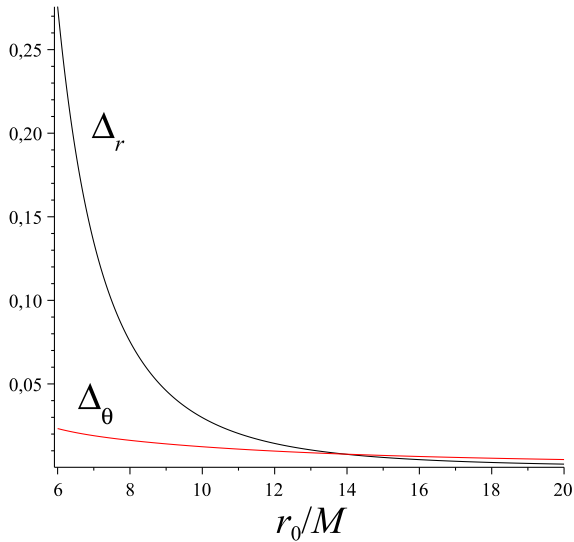}
\end{center}
\caption{The behavior of the variation of the radial distance and polar angle after a full revolution as a function of $r_0$ is shown for the choice of parameters $\sigma_\perp=0.1=\sigma_\parallel$, $\alpha_0=\pi/4$, $\tilde W_{13}(0)=0$, $\tilde X_{11}(0)=0.01$ and $\tilde X_{22}(0)=-0.03$ (so that $K_0\approx-3.7\times10^{-2}$ and $K_4(0)\approx4.3\times10^{-3}$).
The reference circular geodesic is assumed to be co-revolving with respect to increasing values of the azimuthal coordinate.
Color codes are as follows: $\Delta_r$ black, $\Delta_\theta$ red.
}
\label{fig:1}
\end{figure}

\begin{widetext}

\begin{eqnarray}
\label{m2solfin}
m_{(2)}&=&2m_0(M\Omega_{\rm(orb)})^2\left\{
K_4(0)
+\frac{\sigma_\parallel^2}{2\Gamma_K\nu_K^2}\left[
\mp\cos\alpha_0\sin\alpha\sin\Omega_{\rm(orb)}\tau
+\frac{1}{\Gamma_K}\sin\alpha_0\left(\sin\alpha\cos\Omega_{\rm(orb)}\tau-\sin\alpha_0\right)\right.\right.\nonumber\\
&&\left.\left.
+\frac14\Gamma_K(2-3(r_0\zeta_K)^2)(\cos2\alpha-\cos2\alpha_0)
\right]
\right\}\,.
\end{eqnarray} 

Finally, the equations of motion (\ref{papcoreqs1}) give the evolution equations for the second order corrections to the orbit $x^\alpha_{(2)}$, i.e., 
\begin{eqnarray}
\label{eqmoto2}
r_0\frac{\rmd^2 r_{(2)}}{\rmd \tau^2}&=&3N^2(r_0\Omega_{\rm(orb)})^2\frac{r_{(2)}}{r_0}
\pm\frac{2M}{r_0\Omega_{\rm(orb)}}\frac{\rmd \phi_{(2)}}{\rmd \tau}
\pm3(r_0\zeta_K)^3\Gamma_K \left[2U_{(1)}^{\hat \phi}-(r_0\zeta_K)^2\Gamma_K\left(3-7(r_0\zeta_K)^2\right)\sigma_\perp\right]\frac{r_{(1)}}{r_0}\nonumber\\
&&
-2(r_0\zeta_K)^2\Gamma_K^2\left(2-3(r_0\zeta_K)^2\right)\frac{r_{(1)}^2}{r_0^2}
-(r_0\zeta_K)^2\theta_{(1)}^2
+\frac{1}{\Gamma_K^2}\left[(U_{(1)}^{\hat \theta})^2+(U_{(1)}^{\hat \phi})^2\right]\nonumber\\
&&
+3(r_0\zeta_K)^8\left[(\tilde S^{\hat \phi})^2-N^2\Gamma_K^2\sigma_\perp^2\right]
-3(r_0\zeta_K)^4\Gamma_K\left[\left(1+(r_0\zeta_K)^2\right)U_{(1)}^{\hat \phi}\sigma_\perp-N^2U_{(1)}^{\hat \theta}\tilde S^{\hat \phi}\right]\nonumber\\
&&
-2(r_0\zeta_K)^6N^2\Gamma_K^2\left[2(r_0\zeta_K)^2K_1-3K_4\pm2(r_0\zeta_K)r_0\frac{\rmd K_2}{\rmd \tau}\right]
\,,\nonumber\\
r_0^2\frac{\rmd^2 \theta_{(2)}}{\rmd \tau^2}&=&-(r_0\Omega_{\rm(orb)})^2\theta_{(2)}
+2(r_0\Omega_{\rm(orb)})^2\left[\theta_{(1)}\pm\nu_K\left(1-(r_0\zeta_K)^2-3(r_0\zeta_K)^4\right)\tilde S^{\hat r}\right]\frac{r_{(1)}}{r_0}\nonumber\\
&&
\mp2(r_0\Omega_{\rm(orb)})\left[\theta_{(1)}\pm\frac12\nu_K\left(1-(r_0\zeta_K)^2\right)\left(2-3(r_0\zeta_K)^2\right)\tilde S^{\hat r}\right]U_{(1)}^{\hat \phi}\nonumber\\
&&
-2N\left[U_{(1)}^{\hat \theta}+\frac12\Gamma_K\nu_K^2\left(1-(r_0\zeta_K)^2\right)\left(2-9(r_0\zeta_K)^2\right)\tilde S^{\hat \phi}\right]U_{(1)}^{\hat r}
+3N(M\zeta_K)^2\tilde S^{\hat r}\sigma_\perp
\,,\nonumber\\
r_0^2\frac{\rmd^2 \phi_{(2)}}{\rmd \tau^2}&=&\mp2(r_0\Omega_{\rm(orb)})\frac{\rmd r_{(2)}}{\rmd \tau}
-2NU_{(1)}^{\hat r}\left[U_{(1)}^{\hat \phi}\mp2(r_0\Omega_{\rm(orb)})\frac{r_{(1)}}{r_0}\right]
\pm2(r_0\Omega_{\rm(orb)})U_{(1)}^{\hat \theta}\left[\theta_{(1)}\pm\frac32N(r_0\zeta_K)^3\tilde S^{\hat r}\right]\nonumber\\
&&
-3(r_0\zeta_K)^7N\Gamma_K^2\left\{
(r_0\zeta_K)\tilde S^{\hat r}\tilde S^{\hat \phi}
\mp \gamma_K\left[r_0\frac{\rmd K_4}{\rmd \tau}\mp\frac43\frac{1-(r_0\zeta_K)^2}{r_0\zeta_K}K_2\right]
\right\}
\,,
\end{eqnarray} 
with 
\begin{eqnarray}
\frac{\rmd t_{(2)}}{\rmd \tau}&=&\pm\frac{\nu_K^2}{\zeta_K}\frac{\rmd \phi_{(2)}}{\rmd \tau}
+\frac1{N^4}\frac{r_{(1)}}{r_0}\left[\frac32(M\zeta_K)(r_0\Omega_{\rm(orb)})\frac{r_{(1)}}{r_0}\pm \frac{N\nu_K}{\Gamma_K^2}U_{(1)}^{\hat \phi}\right]
-\frac{\Gamma_K}{2N^2}(r_0\zeta_K)^2\theta_{(1)}^2\nonumber\\
&&
+\frac{1}{2N^2\Gamma_K}\left[(U_{(1)}^{\hat r})^2+(U_{(1)}^{\hat \theta})^2+\frac{1}{\gamma_K^2}(U_{(1)}^{\hat \phi})^2\right]
\,,
\end{eqnarray} 
from the normalization condition.

The equation for $\theta_{(2)}$ can be integrated straightforwardly.
The equations for $r_{(2)}$ and $\phi_{(2)}$ are instead coupled. However, taking the derivative of the equation for $r_{(2)}$ with respect to $\tau$ and using the equation for $\phi_{(2)}$ leads to an equation for $U_{(2)}^r$, which can be easily integrated with initial conditions 
\beq
\label{K0def}
U_{(2)}^r(0)=0\,, \qquad
r_0\frac{\rmd U_{(2)}^r(0)}{\rmd \tau}=2(r_0\zeta_K)^6\Gamma_K^2\left[
(2-5(r_0\zeta_K)^2)K_4(0)-\frac32(r_0\zeta_K)^2N^2(K(0)+\sigma_\parallel^2\cos2\alpha_0+\sigma_\perp^2)
\right]\,,
\eeq 
leading to the general solution
\begin{eqnarray}
\label{Ur2sol}
U_{(2)}^r&=&\Omega_{\rm(ep)}(A_0\Omega_{\rm(ep)}\tau-A_1)\cos\Omega_{\rm(ep)}\tau
+\Omega_{\rm(ep)}(A_0-A_2)\sin\Omega_{\rm(ep)}\tau
-2\Omega_{\rm(ep)}A_3\sin2\Omega_{\rm(ep)}\tau
\pm2\zeta_KA_4\sin2\alpha\nonumber\\
&&
+\Omega_{\rm(orb)}\left(\mp\frac{1}{\Gamma_K}A_5-A_6\right)\left[\cos\alpha_0\sin\Omega_{\rm(orb)}\tau\mp\frac{1}{\Gamma_K}\sin\alpha_0\cos\Omega_{\rm(orb)}\tau\right]\cos\alpha\nonumber\\
&&
+\Omega_{\rm(orb)}\left(A_5\pm\frac{1}{\Gamma_K}A_6\right)\left[\cos\alpha_0\cos\Omega_{\rm(orb)}\tau\pm\frac{1}{\Gamma_K}\sin\alpha_0\sin\Omega_{\rm(orb)}\tau\right]\sin\alpha\nonumber\\
&&
+\frac{3(r_0\zeta_K)^2N^2\Gamma_K^2}{r_0\Omega_{\rm(ep)}}\left[
\sin\Omega_{\rm(ep)}\tau\int_0^{\Omega_{\rm(ep)}\tau}\frac{\rmd K}{\rmd\xi}\cos\xi\,\rmd\xi
-\cos\Omega_{\rm(ep)}\tau\int_0^{\Omega_{\rm(ep)}\tau}\frac{\rmd K}{\rmd\xi}\sin\xi\,\rmd\xi
\right]
\,,
\end{eqnarray} 
where 
\beq
\label{Kdef}
K=-\frac23\frac1{(r_0\zeta_K)^2N^2\Gamma_K^4}[\tilde X_{11}-(r_0\Omega_{\rm(orb)})^2\tilde X_{22}]
\eeq
such that 
\beq
K_1=\frac12\nu_K^2\frac{1-(r_0\zeta_K)^2}{(r_0\zeta_K)^4}K_4+\frac34K\,.
\eeq
In order to obtain an explicit solution, we will assume $K=K(0)\equiv K_0=$ const.\footnote{
Alternatively, one can Fourier-expand the quadrupole components entering $K$.
However, we will limit here to the simplest situation $K=K_0=$ constant, which is general enough to capture the main features of the dynamics.}
The solution is then given by
\begin{eqnarray}
\label{solord2}
t_{(2)}&=&\pm \frac{\nu_K^2}{\zeta_K}\phi_{(2)}
+D_1\sin\Omega_{\rm(ep)}\tau
+D_2\sin2\Omega_{\rm(ep)}\tau
+D_3\Omega_{\rm(ep)}\tau
+D_4\sin2\Omega_{\rm(orb)}\tau
+D_5(\cos2\Omega_{\rm(orb)}\tau-1)\nonumber\\
&&
+D_6\left[\cos\alpha_0\cos\alpha\sin\Omega_{\rm(orb)}\tau\mp\frac{1}{\Gamma_K}\sin\alpha_0\left(\cos\alpha\cos\Omega_{\rm(orb)}\tau-\cos\alpha_0\right)\right]
+D_7(\sin2\alpha-\sin2\alpha_0)
\,,\nonumber\\
r_{(2)}&=&\sin\Omega_{\rm(ep)}\tau(A_0\Omega_{\rm(ep)}\tau-A_1)
+A_2(\cos\Omega_{\rm(ep)}\tau-1)
+A_3(\cos2\Omega_{\rm(ep)}\tau-1)
+A_4(\cos2\alpha-\cos2\alpha_0)\nonumber\\
&&
+A_5\left[\cos\alpha_0\sin\alpha\sin\Omega_{\rm(orb)}\tau\mp\frac{1}{\Gamma_K}\sin\alpha_0\left(\sin\alpha\cos\Omega_{\rm(orb)}\tau-\sin\alpha_0\right)\right]\nonumber\\
&&
+A_6\left[\cos\alpha_0\left(\cos\alpha\cos\Omega_{\rm(orb)}\tau-\cos\alpha_0\right)
\pm\frac{1}{\Gamma_K}\sin\alpha_0\cos\alpha\sin\Omega_{\rm(orb)}\tau\right]
\,,\nonumber\\
\theta_{(2)}&=&B_1(\cos\alpha\cos\Omega_{\rm(ep)}\tau-\cos\alpha_0)
+B_2(\cos\alpha-\cos\alpha_0)
+B_3\sin\Omega_{\rm(orb)}\tau
+B_4\sin\alpha\sin\Omega_{\rm(ep)}\tau\nonumber\\
&&
+B_5\left[\cos\alpha_0\sin\Omega_{\rm(orb)}\tau\mp\frac{1}{\Gamma_K}\sin\alpha_0\cos\Omega_{\rm(orb)}\tau\right](\sin\Omega_{\rm(ep)}\tau-\Omega_{\rm(ep)}\tau)
+B_6(\cos\Omega_{\rm(orb)}\tau-1)
\,,\nonumber\\
\phi_{(2)}&=&C_0(\cos\Omega_{\rm(ep)}\tau-1)
+C_1\sin\Omega_{\rm(ep)}\tau
+C_2\sin2\Omega_{\rm(ep)}\tau
+(C_3\cos\Omega_{\rm(ep)}\tau+C_4)\Omega_{\rm(ep)}\tau
+C_5\sin2\Omega_{\rm(orb)}\tau\nonumber\\
&&
+C_6(\cos2\Omega_{\rm(orb)}\tau-1)
+C_7\left[\cos\alpha_0\left(\sin\alpha\cos\Omega_{\rm(orb)}\tau-\sin\alpha_0\right)\pm\frac{1}{\Gamma_K}\sin\alpha_0\sin\alpha\sin\Omega_{\rm(orb)}\tau\right]\nonumber\\
&&
+C_8\left[\cos\alpha_0\cos\alpha\sin\Omega_{\rm(orb)}\tau\mp\frac{1}{\Gamma_K}\sin\alpha_0\left(\cos\alpha\cos\Omega_{\rm(orb)}\tau-\cos\alpha_0\right)
\right]
+C_9(\sin2\alpha-\sin2\alpha_0)
\,.
\end{eqnarray} 
Note that the initial conditions (\ref{inicond1})--(\ref{inicond2}) have been adopted here. A more general choice could be done, introducing additional terms depending on arbitrary constants.   
The explicit expressions for the coefficients are listed in Appendix \ref{const}.
The terms involving the coefficients $A_0$ and $B_5$ are responsible for secular effects in the radial and polar motion respectively, which may lead to observable effects.

Noticeably, in the special case $\sigma_\parallel\equiv0$ corresponding to a spin vector $S=-s_\perp e_{\hat \theta}$ orthogonal to the equatorial plane, the above solution reduces to
\begin{eqnarray}
\label{solord2perp}
t_{(2)}&=&\pm \frac{\nu_K^2}{\zeta_K}\phi_{(2)}
+D_1\sin\Omega_{\rm(ep)}\tau
+D_2\sin2\Omega_{\rm(ep)}\tau
+D_3\Omega_{\rm(ep)}\tau
\,,\nonumber\\
r_{(2)}&=&\sin\Omega_{\rm(ep)}\tau(A_0\Omega_{\rm(ep)}\tau-A_1)
+A_2(\cos\Omega_{\rm(ep)}\tau-1)
+A_3(\cos2\Omega_{\rm(ep)}\tau-1)
\,,\nonumber\\
\phi_{(2)}&=&C_1\sin\Omega_{\rm(ep)}\tau
+C_2\sin2\Omega_{\rm(ep)}\tau
+(C_3\cos\Omega_{\rm(ep)}\tau+C_4)\Omega_{\rm(ep)}\tau
\,,
\end{eqnarray} 
and $\theta_{(2)}=0$, $m_{(2)}=2m_0(M\Omega_{\rm(orb)})^2K_4(0)=$ const., with the limiting expressions for the nonvanishing coefficients easily follow from Eqs. (\ref{solord2coeffsA}), (\ref{solord2coeffsC}) and (\ref{solord2coeffsD}). The motion is confined to the equatorial plane, since $\theta_{(1)}=0$ too.
Furthermore, the quadrupolar quantities $K_2$, $K_3$ and $K_5$ are identically vanishing, whereas $K_1$ and $K_4$ are constant and equal to their initial values.
This particular solution reproduces the results of Ref. \cite{quadrup_schw}.
 
In the case $\sigma_\perp\equiv0$ corresponding to a spin vector oscillating in the equatorial plane, we have, instead, $A_0=0=A_3$, $B_i=0$, $C_2=0=C_3$, $D_1=0=D_2$, implying no secular increase of $r_{(2)}$ during the evolution and $\theta_{(2)}=0$.
Therefore, the motion is oscillating about the reference circular geodesic along both radial and polar directions, due to second order and first order corrections respectively.

One can compute the variation of the radial distance and polar angle after a full revolution, i.e., at the proper time value $\tau=\tau_\ast$ such that $\phi(\tau_\ast)=2\pi$.
We find 
\begin{eqnarray}
\Delta_r\equiv \frac{r(\tau_\ast)-r_0}{r_0}&=&
\mp\epsilon\Sigma_\perp(\cos\xi_0-1)
+\epsilon^2\left\{
-2\Sigma_\perp^2\sin\xi_0(\sin\xi_0-\xi_0)
+\frac1{r_0}\left[A_6\cos\alpha_0(\cos\beta_0-\cos\alpha_0)\right.\right.\nonumber\\
&&\left.\left.
+\sin\xi_0(A_0\xi_0-A_1)
\mp\frac{A_5}{\Gamma_K}\sin\alpha_0(\sin\beta_0-\sin\alpha_0)
+A_2(\cos\xi_0-1)
+A_3(\cos2\xi_0-1)\right.\right.\nonumber\\
&&\left.\left.
+A_4(\cos2\beta_0-\cos2\alpha_0)
\right]
\right\}
\,,\nonumber\\
\Delta_\theta\equiv \frac{\theta(\tau_\ast)-\frac{\pi}{2}}{\frac{\pi}{2}}&=&
\mp\epsilon\frac{2}{\pi}\Sigma_\parallel(\cos\beta_0-\cos\alpha_0)
+\epsilon^2\frac{2}{\pi}\left\{
-\frac{4\pi}{\xi_0\Gamma_K}\Sigma_\parallel\Sigma_\perp(\sin\xi_0-\xi_0)\left[
\sin\alpha_0\left(1-\frac{\xi_0^2}{4\pi^2}\right)-\sin\beta_0
\right]\right.\nonumber\\
&&\left.
+B_1(\cos\beta_0\cos\xi_0-\cos\alpha_0)
+B_2(\cos\beta_0-\cos\alpha_0)
+B_4\sin\beta_0\sin\xi_0
\right\}
\,,
\end{eqnarray} 

\end{widetext}
where
\beq
\xi_0=\pm2\pi\frac{\Omega_{\rm(ep)}}{\Omega_{\rm(orb)}}\,,\qquad
\beta_0=\alpha_0-\frac{2\pi}{\Gamma_K}\,.
\eeq
When the frequencies $\Omega_{\rm(ep)}$ and $\Omega_{\rm(orb)}$ are rationally dependent, the variations $\Delta_r$ and $\Delta_\theta$ get simpler expressions, since $\sin\xi_0=0$ and $\cos\xi_0=1$.
Figure \ref{fig:1} shows a typical behavior of both radial and polar variations for selected values of spin and quadrupole parameters.
They are both monotonically decreasing at large distances, with $\Delta_r$ decreasing faster than $\Delta_\theta$.
Deviations from the geodesic radius are instead dominant at close distances.


Finally, the second order corrections to the circular geodesic conserved energy and angular momentum (\ref{geoEJ}) are given by
\begin{eqnarray}
\label{costmoto2}
E_{(2)}&=&\pm\zeta_K J_{(2)}+2m_0(r_0\zeta_K)^6\Gamma_K K_4(0)\,, \nonumber\\
J_{(2)}&=&\mp m_0r_0(r_0\zeta_K)^5N^2\Gamma_K^3\left[2K_4(0)\right.\nonumber\\
&&\left.
+3(r_0\zeta_K)^2(K_0+\sigma_\perp^2-\sigma_\parallel^2\cos^2\alpha_0)
\right]\,,
\end{eqnarray} 
respectively, as from Eq. (\ref{totalenergy}).

\subsection{Quasi-Keplerian parametrization of the orbit}

It is useful to introduce a Keplerian-like parametrization for the $r$-$\phi$ motion \cite{dam-derue1,dam-derue2,dam88,wex}, i.e.,
\begin{eqnarray}
\label{quasikeplgen}
\frac{2\pi}{P}(t-t_0)&=& \ell_t -e_t \sin \ell_t\,,\nonumber\\
r&=&a_r(1-e_r \cos \ell_r)\,, \nonumber\\
\frac{2\pi}{\Phi}(\phi-\phi_0)&=& 2\arctan\left(\sqrt{\frac{1+e_\phi}{1-e_\phi}}\tan\frac{\ell_\phi}{2}\right)
\,.
\end{eqnarray}
The quantities $e_t$, $e_r$ and $e_\phi$ are three eccentricities, while $P$ and $\Phi$ denote the periods of $t$ and $\phi$ motions, respectively (with an abuse of notation for $P$, not to be confused with the body's 4-momentum).
The quantities $\ell_t$, $\ell_r$ and $\ell_\phi$ are functions of the proper time parameter $\tau$ on the orbit.
We find
\begin{eqnarray}
\ell_t&=&\ell+\epsilon^2\ell_t^{(2)}\,,\nonumber\\
\ell_r&=&\ell+\epsilon\ell_r^{(1)}\,,\nonumber\\
\ell_\phi&=&\ell+\epsilon^2\ell_\phi^{(2)}\,,
\end{eqnarray}
where $\ell =\Omega_{\rm(ep)}\tau$ and 
\begin{eqnarray}
\ell_t^{(2)}&=&
-\frac{\Omega_{\rm (ep)}}{\Gamma_K}\left[t_{(2)}(\ell)-\left(D_3\pm\frac{\nu_K^2}{\zeta_K}C_4\right)\ell\right.\nonumber\\
&&\left.
-\left(D_1\pm\frac{\nu_K^2}{\zeta_K}C_1\right)\sin\ell\right]
\,,\nonumber\\
\ell_r^{(1)}&=&\pm\frac{r_{(2)}(\ell)-A_2(\cos\ell-1)+2A_3}{r_0\Sigma_\perp\sin\ell}
\,,\nonumber\\
\ell_\phi^{(2)}&=&
\pm\frac{\Omega_{\rm (ep)}}{\Omega_{\rm (orb)}}\left[\phi_{(2)}(\ell)-C_4\ell-C_1\sin\ell\right.\nonumber\\
&&\left.
\mp\frac{\Omega_{\rm (orb)}}{\Omega_{\rm (ep)}}\Sigma_\perp^2\sin2\ell
\right]\,.
\end{eqnarray}

The semi-major axis and the eccentricities turn out to be
\beq
\label{solarquad}
a_r = r_0\pm\epsilon r_0\Sigma_\perp-\epsilon^2(A_2+2A_3)\,,
\eeq
and 
\begin{eqnarray}
\label{soleccquad}
e_t &=&\mp2\epsilon\nu_K^2\Sigma_\perp\nonumber\\
&&
-\epsilon^2\frac{\Omega_{\rm (ep)}}{\Gamma_K}\left[D_1\pm\frac{\nu_K^2}{\zeta_K}\left(C_1\pm4\nu_K^2\frac{\Omega_{\rm (orb)}}{\Omega_{\rm (ep)}}\Sigma_\perp^2\right)\right]
\,, \nonumber\\
e_r &=&\pm\epsilon\Sigma_\perp-\epsilon^2\left(\frac{A_2}{r_0}+\Sigma_\perp^2\right)
\,,\nonumber\\
e_\phi &=&\pm2\epsilon \Sigma_\perp\pm\epsilon^2\left(\frac{\Omega_{\rm (ep)}}{\Omega_{\rm (orb)}}C_1\pm4\Sigma_\perp^2\right)\,,
\end{eqnarray}
respectively, whereas the periods of $t$ and $\phi$ motions read
\begin{eqnarray}
\label{solper1quad}
\frac{P}{2\pi}&=&\frac{\Gamma_K}{\Omega_{\rm (ep)}}\left[1\mp2\epsilon\nu_K^2\Sigma_\perp
+\epsilon^2\frac{\Omega_{\rm (ep)}}{\Gamma_K}\left(D_3\pm\frac{\nu_K^2}{\zeta_K}C_4\right)\right]\,, \nonumber\\
\frac{\Phi}{2\pi}&=& \pm\frac{\Omega_{\rm (orb)}}{\Omega_{\rm (ep)}}\left[1\mp2\epsilon\Sigma_\perp\pm\epsilon^2\frac{\Omega_{\rm (ep)}}{\Omega_{\rm (orb)}}C_4\right]
\,.
\end{eqnarray}
Note that the first order corrections to the orbital parameters only depend on $\sigma_\perp$, because the spin precession in the $r$-$\phi$ plane affects only the $\theta$-motion to that order.

The interest of such quantities is that they are directly related to observable effects. Indeed, one can measure in principle both the orbital period and the fractional periastron advance \cite{dam88} defined by 
\beq
\label{peradv}
k\equiv\frac{\Phi}{2\pi}-1\,.
\eeq

In the weak field limit $u_0\equiv M/r_0\ll1$ the above expressions become

\begin{widetext}

\begin{eqnarray}
e_t^{\rm wf} &=&
\mp6\sigma_\perp u_0^{5/2}(1+6u_0)\epsilon
-u_0^3\left[\frac23(3-5\cos2\alpha_0)\sigma_\parallel^2+12\tilde X_{11}(0)\pm32\tilde W_{13}(0)u_0^{1/2}\right]\epsilon^2
+O(u_0^4)
\,, \nonumber\\
e_r^{\rm wf} &=&
\pm3\sigma_\perp u_0^{3/2}(1+4u_0+24u_0^2)\epsilon
+u_0^2\left\{
\frac13(3-5\cos2\alpha_0)\sigma_\parallel^2+6\tilde X_{11}(0)
\pm8\tilde W_{13}(0)u_0^{1/2}(2+5u_0)\right.\nonumber\\
&&\left.
+\left[10\tilde X_{11}(0)+2\tilde X_{22}(0)+\frac13(13+5\cos2\alpha_0)\sigma_\parallel^2-12\sigma_\perp^2\right]u_0
\right\}\epsilon^2
+O(u_0^4)
\,,\nonumber\\
e_\phi^{\rm wf} &=&
2e_r^{\rm wf}+18\epsilon^2u_0^3\sigma_\perp^2
+O(u_0^4)
\,,\nonumber\\
\frac{a_r^{\rm wf}}{r_0} &=&1+e_r^{\rm wf}+18\epsilon^2u_0^3\sigma_\perp^2
+O(u_0^4)
\,, 
\end{eqnarray}
and
\begin{eqnarray}
\frac1{M}\frac{P^{\rm wf}}{2\pi}&\equiv&\frac{1}{2\pi M}\left(P_{(0)}^{\rm wf}+\epsilon P_{(1)}^{\rm wf}+\epsilon^2 P_{(2)}^{\rm wf}\right)\nonumber\\
&=&u_0^{-3/2}\left(1+3u_0+\frac{27}{2}u_0^2+\frac{135}{2}u_0^3+\frac{2835}{8}u_0^4+\frac{15309}{8}u_0^5\right)
\mp3u_0(2+18u_0+135u_0^2)\sigma_\perp\epsilon\nonumber\\
&&
+u_0^{1/2}\left\{2\sigma_\parallel^2
+\left[(10+3\cos2\alpha_0)\sigma_\parallel^2-12\tilde X_{11}(0)\right]u_0
\mp16\tilde W_{13}(0)u_0^{3/2}(2+15u_0)\right.\nonumber\\
&&\left.
+\left[\frac{69}{4}(4+\cos2\alpha_0)\sigma_\parallel^2+33\sigma_\perp^2-80\tilde X_{11}(0)-4\tilde X_{22}(0)\right]u_0^2\right.\nonumber\\
&&\left.
+\left[\frac{9}{2}(102+25\cos2\alpha_0)\sigma_\parallel^2+504\sigma_\perp^2-558\tilde X_{11}(0)-36\tilde X_{22}(0)\right]u_0^3
\right\}\epsilon^2
+O(u_0^4)
\,, \nonumber\\
\frac{\Phi^{\rm wf}}{2\pi}&\equiv&\frac{1}{2\pi}\left(\Phi_{(0)}^{\rm wf}+\epsilon \Phi_{(1)}^{\rm wf}+\epsilon^2 \Phi_{(2)}^{\rm wf}\right)\nonumber\\
&=&
\pm\left(1+3u_0+\frac{27}{2}u_0^2+\frac{135}{2}u_0^3\right)
-3\sigma_\perp u_0^{3/2}(2+14u_0+99u_0^2)\epsilon\nonumber\\
&&
\mp u_0\left\{2\sigma_\parallel^2
+\left[\frac34(9+4\cos2\alpha_0)\sigma_\parallel^2-12\tilde X_{11}(0)\right]u_0
\mp16\tilde W_{13}(0)u_0^{3/2}(2+11u_0)\right.\nonumber\\
&&\left.
+\left[\frac14(199+45\cos2\alpha_0)\sigma_\parallel^2+\frac{75}{2}\sigma_\perp^2-56\tilde X_{11}(0)-4\tilde X_{22}(0)\right]u_0^2
\right\}\epsilon^2
+O(u_0^4)
\,.
\end{eqnarray}

\end{widetext}
Simple inspection of the above formulas shows that the second order corrections to both periods involve the square of the spin parameters $\sigma_\perp$ and $\sigma_\parallel$ as well as the quadrupole parameters $\tilde X_{11}(0)$, $\tilde X_{22}(0)$ and $\tilde W_{13}(0)$: the relative weight of $\sigma_\parallel$ to $\sigma_\perp$ contributions behaves as $u_0^2$, whereas that of $\sigma_\parallel$ to $\tilde X_{11}(0)$, $\tilde X_{22}(0)$ and $\tilde W_{13}(0)$ as $u_0$ and $u_0^{3/2}$, respectively, implying an actual dominant role for $\sigma_\parallel$.

\subsection{Evolution of the quadrupolar structure of the body}

Let us study now the evolution of the quadrupolar structure of the body as given by Eqs. (\ref{compatib}). 
They read

\begin{widetext}

\begin{eqnarray}
\label{compatib2}
K_1&=&\frac12\nu_K^2\frac{1-(r_0\zeta_K)^2}{(r_0\zeta_K)^4}K_4+\frac34K
\,,\nonumber\\
K_2&=&\frac14\frac{\sigma_\parallel^2}{(r_0\zeta_K)^2}\left\{
\left(1-\frac32(r_0\zeta_K)^2\right)\sin2\alpha
\mp\frac{1}{\Gamma_K}\sin\alpha_0\left(\sin\alpha\sin\Omega_{\rm(orb)}\tau\pm\frac{1}{\Gamma_K}\cos\alpha\cos\Omega_{\rm(orb)}\tau\right)\right.\nonumber\\
&&\left.
-\cos\alpha_0\left(\sin\alpha\cos\Omega_{\rm(orb)}\tau\mp\frac{1}{\Gamma_K}\cos\alpha\sin\Omega_{\rm(orb)}\tau\right)
\right\}
\,,\nonumber\\
K_3&=&\frac14\frac{\sigma_\parallel\sigma_\perp}{(r_0\zeta_K)^4}\nu_K\left\{
\frac{1}{\Gamma_K^3}\left(\cos\alpha_0\sin\Omega_{\rm(orb)}\tau\mp\frac{1}{\Gamma_K}\sin\alpha_0\cos\Omega_{\rm(orb)}\tau\right)
-3\frac{\Omega_{\rm(orb)}}{\Omega_{\rm(ep)}}\frac{(r_0\zeta_K)^4}{\Gamma_K}\cos\alpha\sin\Omega_{\rm(ep)}\tau\right.\nonumber\\
&&\left.
\pm\frac{\Omega_{\rm(orb)}^2}{\Omega_{\rm(ep)}^2}\sin\alpha\left[
1-18(r_0\zeta_K)^2+81(r_0\zeta_K)^4-99(r_0\zeta_K)^6
+3(r_0\zeta_K)^2(2-12(r_0\zeta_K)^2+15(r_0\zeta_K)^4)\cos\Omega_{\rm(ep)}\tau
\right]\right\}
\,,\nonumber\\
K_4&=&
\frac1{2(M\Omega_{\rm(orb)})^2}\frac{m_{(2)}}{m_0}
\,,\nonumber\\
K_5&=&\pm\frac14\frac{\sigma_\parallel\sigma_\perp}{(r_0\zeta_K)^4}\nu_K\left\{
\frac{1}{\Gamma_K^2}\left(\cos\alpha_0\cos\Omega_{\rm(orb)}\tau\pm\frac{1}{\Gamma_K}\sin\alpha_0\sin\Omega_{\rm(orb)}\tau\right)\right.\nonumber\\
&&\left.
-\frac{\Omega_{\rm(orb)}^2}{\Omega_{\rm(ep)}^2}\cos\alpha\left[
1-3(r_0\zeta_K)^2-9(r_0\zeta_K)^4+9(r_0\zeta_K)^6
-3\frac{1}{\Gamma_K^2}(r_0\zeta_K)^2(2-5(r_0\zeta_K)^2)\cos\Omega_{\rm(ep)}\tau
\right]\right\}
\,,
\end{eqnarray} 

\end{widetext}
where $m_{(2)}$ is given by Eq. (\ref{m2solfin}).

\begin{figure*} 
\typeout{*** EPS figure 2}
\begin{center}
$\begin{array}{cc}
\includegraphics[scale=0.4]{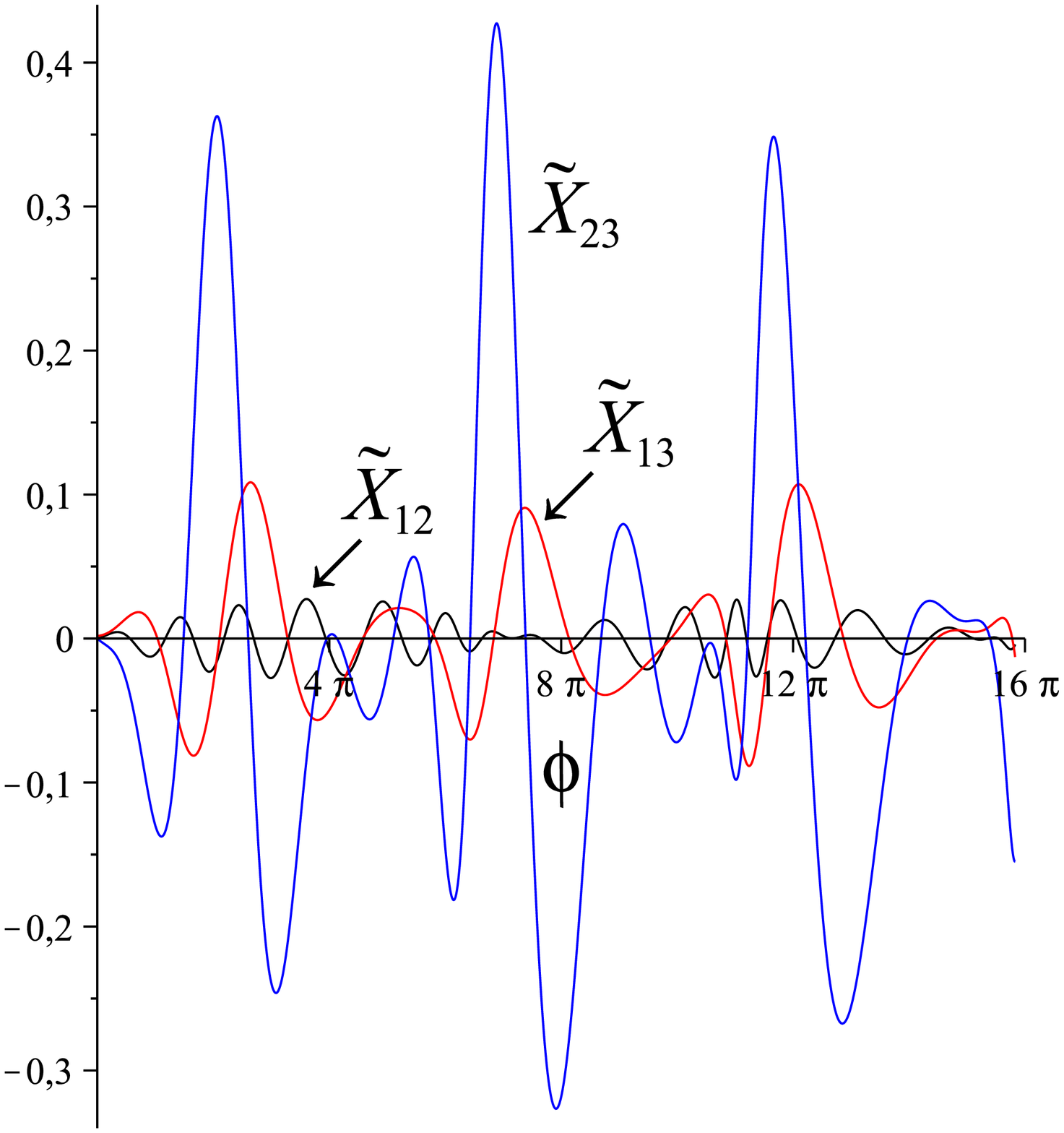}&\quad
\includegraphics[scale=0.42]{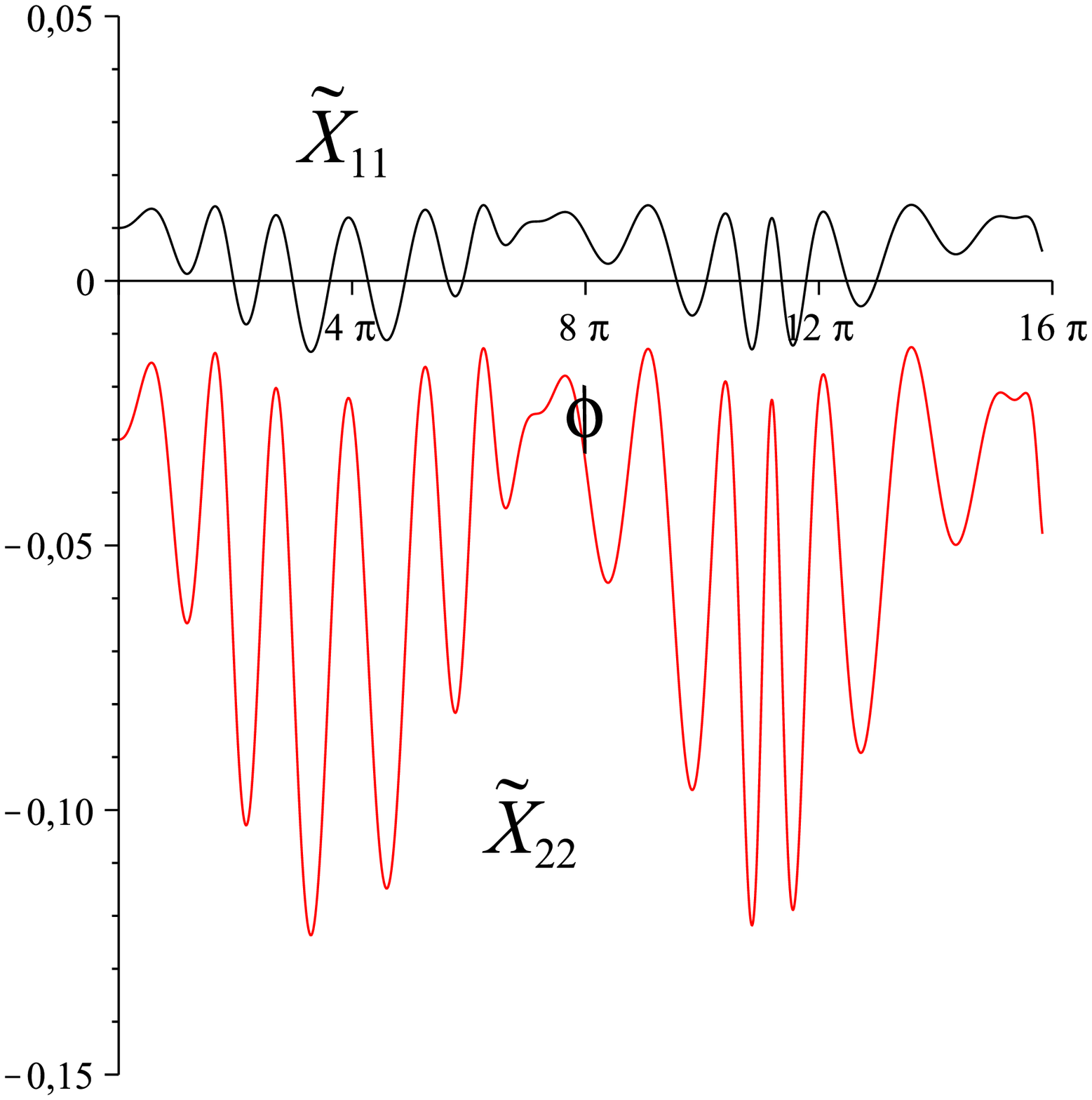}\\[.4cm]
\quad\mbox{(a)}\quad &\quad \mbox{(b)}
\end{array}$\\
\end{center}
\caption{The behavior of the (dimensionless) frame components $\tilde X_{ab}$ of the quadrupole tensor as a function of the azimuthal angle $\phi$ along the orbit of the extended body is shown for the choice of parameters $r_0/M=7$, $\sigma_\perp=0.1=\sigma_\parallel$, $\alpha_0=\pi/4$, $\tilde X_{11}(0)=0.01$ and $\tilde X_{22}(0)=-0.03$ (so that $K_0\approx-3.7\times10^{-2}$ and $K_4(0)\approx4.3\times10^{-3}$). The reference circular geodesic is assumed to be co-revolving.
Color codes are as follows: $\tilde X_{12}$ black, $\tilde X_{13}$ red and $\tilde X_{23}$ blue in panel (a); $\tilde X_{11}$ black and $\tilde X_{22}$ red in panel (b).
}
\label{fig:2}
\end{figure*}

Let us neglect the magnetic components of the quadrupole tensor (i.e., $\tilde W_{ab}=0$), for simplicity.
Therefore, the quadrupolar structure of the body is completely determined by the mass quadrupole moment spatial tensor $X(u)$, whose frame components vary along the orbit according to Eqs. (\ref{Kdefs}) and (\ref{compatib2}). 
Notice that their evolution is governed by the first order solution only.
However, it is interesting to consider their variation after each revolution of the extended body around the central source, thus requiring the knowledge of the second order solution obtained above.
Hence, we assume the components $\tilde X_{11}$ and $\tilde X_{22}$ to be constrained by the relation (\ref{Kdef}) with $K=K_0=$ const.
Since the initial values of the non-diagonal components are all vanishing, one can then form the electric part of the quadrupole tensor as depending on two parameters only, i.e., $\tilde X_{11}(0)$ and $\tilde X_{22}(0)$ (or, equivalently, $K_0$ and $K_4(0)$).
The behavior of the frame components $\tilde X_{ab}$ as a function of the azimuthal angle $\phi$ along the orbit of the extended body is shown in Fig. \ref{fig:2} for selected values of the orbital and spin parameters.

\begin{figure} 
\typeout{*** EPS figure 3}
\begin{center}
\includegraphics[scale=0.4]{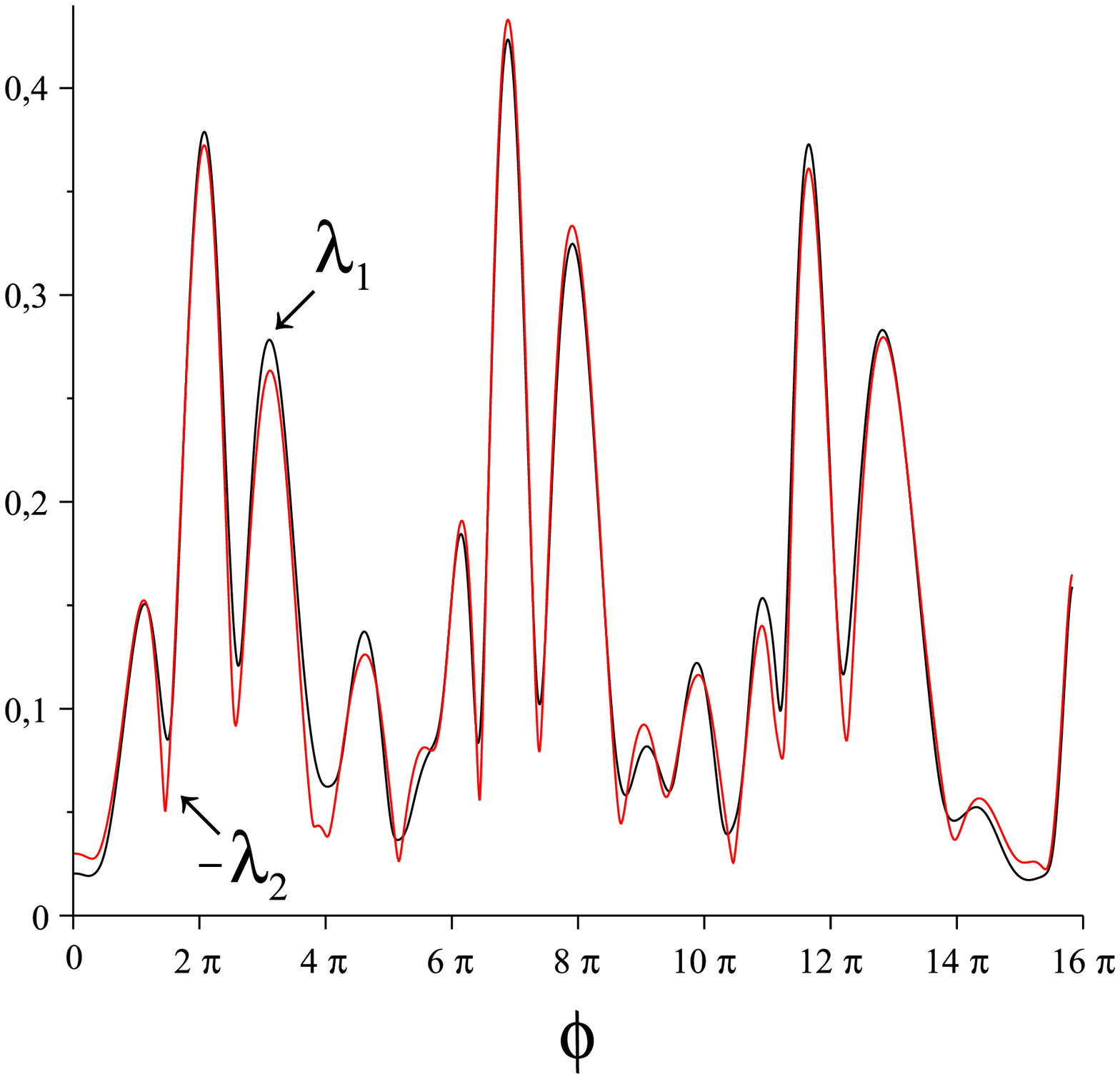}
\end{center}
\caption{The behavior of the eigenvalues $\lambda_1$ and $-\lambda_2$ as a function of the azimuthal angle $\phi$ along the orbit is shown for the same choice of parameters as in Fig. \ref{fig:2}.
Color codes are as follows: $\lambda_1$ black and $-\lambda_2$ red.
}
\label{fig:3}
\end{figure}

Ehlers and Rudolph \cite{ehlers77} and Dixon \cite{dixon_varenna} proposed a relativistic generalization of the moment of inertia tensor of the body, directly relating the moment of inertia to the mass quadrupole tensor as for the familiar quantities of Newtonian mechanics.
The trace-free part of the the moment of inertia tensor can then be identified with the mass quadrupole moment spatial tensor $X(u)$.
One can also introduce a special triad $\{E_{\hat a}\}$ adapted to $u$, which represents the relativistic analogue of the Newtonian \lq\lq body-fixed" spatial frame. The latter can be conveniently chosen so that the mass quadrupole moment spatial tensor  be diagonal in such a frame, i.e., the frame vectors $E_{\hat a}$ are eigenvectors of $X(u)$ and can be identified with the principal axes of the body.
The properties of the associated eigenvalues thus provide information about deviation from sphericity of the shape of the body, which varies with time.
The spatial tensor $\tilde X$ has zero trace, but ${\rm Tr}[\tilde X^2]$ and ${\rm Tr}[\tilde X^3]$ are nonzero, so that the eigenvalue equation results in 
\beq
\lambda^3-\frac12{\rm Tr}[\tilde X^2]\lambda-\frac13{\rm Tr}[\tilde X^3]=0\,,
\eeq 
with solutions $\lambda_1$, $\lambda_2$ and $-(\lambda_1+\lambda_2)$, which can be numerically studied during the evolution.
The behavior of the eigenvalues $\lambda_1$ and $-\lambda_2$ as a function of $\tau$ is shown in Fig. \ref{fig:3} for the same choice of parameters as in Fig. \ref{fig:2}.
The shape of the body significantly changes during the evolution, becoming approximately spherical for those values of $\tau$ such that $|\lambda_{1,2}|\ll1$.


\subsection{Circular orbits}

Finally, it is interesting to consider the special case in which the orbit of the extended body remains circular, which has to be treated separately, due to the choice of initial conditions (\ref{inicond2}).
This implies that the motion be confined to the equatorial plane with spin vector orthogonal to it, i.e., $\sigma_\parallel=0$. 
The associated 4-velocity is 
\beq
\label{casocircU}
U=\Gamma(\partial_t+\zeta\partial_\phi)\,,
\eeq
with normalization factor
\beq
\Gamma=\Gamma_K\left[1\mp \frac32\sigma_\perp(r_0\zeta_K)^5\Gamma_K^2\epsilon+\epsilon^2\tilde\Gamma_{(2)}\right]\,,
\eeq
and angular velocity  
\beq
\label{zetaspin}
\zeta=\pm\zeta_K\left[1\mp \frac32\sigma_\perp(r_0\zeta_K)^3\epsilon+\epsilon^2\tilde\zeta_{(2)}\right]\,,
\eeq
where
\begin{eqnarray}
\tilde\Gamma_{(2)}&=&
(r_0\zeta_K)^2\Gamma_K^2\left[
\tilde\zeta_{(2)}+\frac98(r_0\zeta_K)^6\Gamma_K^2\sigma_\perp^2
\right]
\,,\nonumber\\
\tilde\zeta_{(2)}&=&
(r_0\zeta_K)^4\left[
\frac32(r_0\zeta_K)^2\left(K_0+\frac74\sigma_\perp^2\right)\right.\nonumber\\
&&\left.
-\frac1{N^2}(2-5(r_0\zeta_K)^2)K_4(0)
\right]
\,.
\end{eqnarray} 
The quadrupolar quantities (\ref{Kdefs}) turn out to be either identically vanishing ($K_2=0=K_3=K_5$) or constant ($K_1=K_1(0)$, $K_4=K_4(0)$ and $K=K(0)\equiv K_0$), implying that $\tilde X_{11}$, $\tilde X_{22}$ and $\tilde W_{13}$ are constant and equal to their initial values.
The remaining frame components of the quadrupole tensor, instead, must satisfy the following conditions
\begin{eqnarray}
0&=&\tilde X_{12}\pm2\nu_K\tilde W_{23}
\,,\nonumber\\
0&=&2\tilde W_{12}\mp\nu_K\tilde X_{23}
\,,\nonumber\\
0&=&2(2\tilde W_{11}+\tilde W_{22})\mp\nu_K\frac{1-(r_0\zeta_K)^2}{(r_0\zeta_K)^2}\tilde X_{13}
\,.
\end{eqnarray} 

The parametric equations of the orbit are 
\beq
\label{casocircorb}
t=t_0+\Gamma \tau\,,\quad 
r=r_0\,,\quad 
\theta=\frac{\pi}{2}\,,\quad
\phi=\phi_0+\Omega \tau\,,
\eeq
with orbital angular velocity $\Omega=\Gamma\zeta$.
The relation between $U$ and the unit timelike vector $u$ aligned with the 4-momentum is given by
\beq
u-U=-3\epsilon^2(r_0\zeta_K)^7N\Gamma_K^2\left[\frac43K_1(0)+\sigma_\perp^2\right]{\bar U}_K\,.
\eeq 

Finally, the first and second order corrections to the circular geodesic conserved energy and angular momentum (\ref{geoEJ}) are given by
\begin{eqnarray}
\label{costmoto1circ}
E_{(1)}&=&\pm\frac12 m_0(r_0\zeta_K)^5\Gamma_K^3\sigma_\perp\,, \nonumber\\
E_{(1)}\mp\zeta_K J_{(1)}&=&\mp m_0\frac{(r_0\zeta_K)^3}{\Gamma_K}\sigma_\perp\,,
\end{eqnarray} 
and
\begin{eqnarray}
\label{costmoto2circ}
E_{(2)}&=&-m_0(r_0\zeta_K)^6\Gamma_K^3\left[\frac32(r_0\zeta_K)^2\gamma_K^2K_0\right.\nonumber\\
&&\left.
+(2-3(r_0\zeta_K)^2)K_4(0)\right.\nonumber\\
&&\left.
+\frac38(r_0\zeta_K)^2\gamma_K^2(2-15(r_0\zeta_K)^2)\sigma_\perp^2
\right]\,, \nonumber\\
E_{(2)}\mp\zeta_K J_{(2)}&=&\frac18m_0(r_0\zeta_K)^6\Gamma_K\left[16K_4(0)\right.\nonumber\\
&&\left.
+9(r_0\zeta_K)^2\gamma_K^2\sigma_\perp^2
\right]\,,
\end{eqnarray} 
respectively, as from Eq. (\ref{totalenergy}).

These results generalize those of Ref. \cite{quadrup_schw}, where the quadrupole tensor was assumed to be of purely electric type and with vanishing non-diagonal frame components.

\section{Concluding remarks}

We have studied the dynamics of an extended body with arbitrary spin orientation and quadrupolar structure in a Schwarzschild spacetime within the framework of the Mathisson-Papapetrou-Dixon model.
The equations of motion have been solved analytically in the limit of small values of the characteristic length scales associated with the spin and quadrupole variables with respect to the characteristic length of the background curvature. 
The solution provides the corrections to the circular geodesic motion on the equatorial plane (taken as the reference trajectory) due to both the spin-curvature and quadrupole-curvature couplings.
Terms which are linear in spin are referred to as \lq\lq first order,'' whereas terms quadratic in spin as well as linear in the quadrupole as \lq\lq second order.'' 

The solution to the first order, i.e., for a purely spinning particle, was already known.
The component of the spin vector orthogonal to the unperturbed orbital reference plane remains constant, whereas the component in the equatorial plane undergoes precession, due to the evolution equations. 
The latter component is responsible for the oscillation of the body path about the equatorial plane.
Deviations from the reference circular geodesic are thus governed by three fundamental frequencies: orbital, epicyclic and spin precession frequency. The radial and polar motions are decoupled and exhibit periodic oscillations around the reference geodesic (simply periodic in the radial direction).

The contribution due to the quadrupole is taken into account in the MPD model by including an additional force in the equations of motion and a torque term in the evolution equations for the spin tensor.
We have relaxed most of the simplifying assumptions commonly adopted in the literature concerning both the structure of the body and its dynamics, leading to special solutions: spin vector with constant magnitude and orthogonal to the orbital plane, quadrupole tensor either identically vanishing or with constant components or even fully determined by the spin structure itself (spin-induced quadrupole).
This overcomes the problem of determining the evolution of the quadrupole tensor components along the orbit of the extended body, due to the lack of evolution equations for the quadrupole and higher multipoles in the MPD model. But, on the other hand, under such restrictions the description of the system is limited to very special, highly symmetric situations.

We have considered here the most general form of the quadrupole tensor, which is completely specified by two symmetric and trace-free spatial tensors, representing the mass quadrupole moment (of electric type) and the current quadrupole moment (of magnetic type), each of them having five independent components.
Such ten components are arbitrary in principle. 
However, we have shown that they are constrained by some compatibility conditions as a consequence of the spin evolution equations, which also involve spin vector components and first order corrections to the orbit, whose time variation is determined by the first order solution.  
We have then computed the second order solution for the orbit and studied the associated dynamical observables, including periastron advance, orbital period, eccentricities, etc. The presence of the quadrupole in its full generality leads to multi-frequency coupled spatial motions, exhibiting in addition secular drifts which vary with spin orientation.

Finally, we have recalled the notion of relativistic moment of inertia tensor of the body, which is related to the mass quadrupole moment tensor in the usual way as their Newtonian counterparts in the Newtonian limit do. This formal analogy allows to study how the shape of the body changes along the path by computing the components of the mass quadrupole moment spatial tensor (or, equivalently, the inertia tensor) with respect to a \lq\lq body-fixed" spatial frame.  
We have shown that the shape of the body may undergo significant variations during the evolution, passing from nearly spherical to highly deformed configurations.
This is a purely gravitational effect which may contribute to the variation of the orbital elements of astrophysical systems, like binary pulsar systems.

\acknowledgments
The authors acknowledge ICRANet and the INFN Section of Naples for partial support.

\appendix 

\begin{widetext} 

\section{Second order solution coefficients}
\label{const}

We list below the integration constants of the second order solution (\ref{solord2}):
%
%
%
\begin{eqnarray} 
\label{solord2coeffsA}
A_0&=&
-9MN^4u_0^2\frac{\Omega_{\rm(orb)}^6}{\Omega_{\rm(ep)}^6}(1-9u_0)\sigma_\perp^2
\,,\nonumber\\
A_1&=&
\mp\frac16Mu_0\Gamma_K\frac{\Omega_{\rm(orb)}}{\Omega_{\rm(ep)}}\frac{8-103u_0+435u_0^2-531u_0^3-81u_0^4}{1-6u_0-3u_0^2}\sigma_\parallel^2\sin2\alpha_0
\,,\nonumber\\
A_2&=&
Mu_0\frac{\Omega_{\rm(orb)}^2}{\Omega_{\rm(ep)}^2}\left\{
3u_0N^2K_0-2(2-5u_0)K_4(0)
+3u_0N^2\frac{\Omega_{\rm(orb)}^4}{\Omega_{\rm(ep)}^4}(1+12u_0^2)\sigma_\perp^2\right.\nonumber\\
&&\left.
-\left[
N^2\frac{1-5u_0+3u_0^2-9u_0^3}{1-6u_0-3u_0^2}
-\frac13\frac{5-73u_0+324u_0^2-405u_0^3-81u_0^4}{1-6u_0-3u_0^2}\cos2\alpha_0
\right]\sigma_\parallel^2
\right\}
\,,\nonumber\\
A_3&=&
-\frac92MN^4u_0^2\frac{\Omega_{\rm(orb)}^6}{\Omega_{\rm(ep)}^6}(1-7u_0)\sigma_\perp^2
\,,\nonumber\\
A_4&=&
-\frac{M}{12}(4-2u_0-21u_0^2)\sigma_\parallel^2
\,,\nonumber\\
A_5&=&
\mp\frac23MN^2\Gamma_K\frac{1-2u_0+9u_0^3}{1-6u_0-3u_0^2}\sigma_\parallel^2
\,,\nonumber\\
A_6&=&
\frac23M\frac{N^2}{\Gamma_K^2}\frac{2-u_0}{1-6u_0-3u_0^2}\sigma_\parallel^2
\,,
\end{eqnarray} 
%
%
%
\begin{eqnarray} 
\label{solord2coeffsB}
B_1&=&
-3Nu_0^3\frac{\Omega_{\rm(orb)}^2}{\Omega_{\rm(ep)}^2}\sigma_\perp\sigma_\parallel
\,,\nonumber\\
B_2&=&
\frac{Nu_0}{\Gamma_K^2}\frac{\Omega_{\rm(orb)}^2}{\Omega_{\rm(ep)}^2}(1-5u_0)\sigma_\perp\sigma_\parallel
\,,\nonumber\\
B_3&=&
\mp u_0N\Gamma_K(1-5u_0+9u_0^2)\sigma_\perp\sigma_\parallel\sin\alpha_0
\,,\nonumber\\
B_4&=&
\pm3u_0^3N\Gamma_K\frac{\Omega_{\rm(orb)}}{\Omega_{\rm(ep)}}\sigma_\perp\sigma_\parallel
\,,\nonumber\\
B_5&=&
-6N^3u_0^2\frac{\Omega_{\rm(orb)}}{\Omega_{\rm(ep)}}\sigma_\perp\sigma_\parallel
\,,\nonumber\\
B_6&=&
-u_0N^3\sigma_\perp\sigma_\parallel\cos\alpha_0
\,,\nonumber\\
\end{eqnarray}
%
%
%
\begin{eqnarray} 
\label{solord2coeffsC}
C_0&=&
\mp2\frac{u_0}{M}\frac{\Omega_{\rm(orb)}}{\Omega_{\rm(ep)}}A_1
\,,\nonumber\\
C_1&=&
\mp2\frac{u_0}{M}\frac{\Omega_{\rm(orb)}}{\Omega_{\rm(ep)}}\left[
A_2+9Mu_0^2N^2\frac{\Omega_{\rm(orb)}^6}{\Omega_{\rm(ep)}^6}\sigma_\perp^2
\right]
\,,\nonumber\\
C_2&=&
\pm\frac94N^4u_0^3\frac{\Omega_{\rm(orb)}^7}{\Omega_{\rm(ep)}^7}(5-32u_0)\sigma_\perp^2
\,,\nonumber\\
C_3&=&
\pm2\frac{u_0}{M}\frac{\Omega_{\rm(orb)}}{\Omega_{\rm(ep)}}A_0\,,\nonumber\\
C_4&=&
\pm u_0\frac{\Omega_{\rm(orb)}^3}{\Omega_{\rm(ep)}^3}\left\{
6u_0^2N^2K_0-4u_0(2-5u_0)K_4(0)
+\frac32u_0^2N^2\frac{\Omega_{\rm(orb)}^4}{\Omega_{\rm(ep)}^4}(25-162u_0+288u_0^2)\sigma_\perp^2\right.\nonumber\\
&&\left.
+\frac{N^2}{4}\left[
8-29u_0+6u_0^2
+3u_0\Gamma_K^2(1-4u_0)(4-9u_0)\cos2\alpha_0
\right]\sigma_\parallel^2
\right\}
\,,\nonumber\\
C_5&=&
\pm\frac18N^2u_0\left[(2-3u_0)\cos2\alpha_0+3u_0\right]\sigma_\parallel^2
\,,\nonumber\\
C_6&=&
-\frac14N^2\frac{u_0}{\Gamma_K}\sigma_\parallel^2\sin2\alpha_0
\,,\nonumber\\
C_7&=&
-\frac43N^2\Gamma_K\frac{1-5u_0+8u_0^2}{1-6u_0-3u_0^2}\sigma_\parallel^2
\,,\nonumber\\
C_8&=&
\mp\frac13N^2\frac{1-9u_0-14u_0^2+3u_0^3}{1-6u_0-3u_0^2}\sigma_\parallel^2
\,,\nonumber\\
C_9&=&
-\frac{1}{24}u_0\Gamma_K^3(14-67u_0+42u_0^2+90u_0^3)\sigma_\parallel^2
\,,
\end{eqnarray}
%
%
%
\begin{eqnarray} 
\label{solord2coeffsD}
D_1&=&
9Mu_0^{5/2}\frac{\Omega_{\rm(orb)}^3}{\Omega_{\rm(ep)}^3}\sigma_\perp^2
=-4D_2
\,,\nonumber\\
D_3&=&
-\frac34Mu_0^{3/2}\frac{\Omega_{\rm(orb)}}{{\Omega_{\rm(ep)}}}\left[\sigma_\parallel^2+6u_0\frac{\Omega_{\rm(orb)}^2}{\Omega_{\rm(ep)}^2}\sigma_\perp^2\right]
\,,\nonumber\\
D_4&=&
-\frac18Mu_0^{1/2}\left[(2-3u_0)\cos2\alpha_0+3u_0\right]\sigma_\parallel^2
=\mp\frac{Mu_0^{-1/2}}{N^2}C_5
\,,\nonumber\\
D_5&=&
\pm\frac14M\frac{u_0^{1/2}}{\Gamma_K}\sigma_\parallel^2\sin2\alpha_0
\,,\nonumber\\
D_6&=&
Mu_0^{1/2}\sigma_\parallel^2
\,,\nonumber\\
D_7&=&
\pm\frac18Mu_0^{1/2}\Gamma_K(2-3u_0)\sigma_\parallel^2
\,,
\end{eqnarray} 
where
\begin{eqnarray} 
N=\sqrt{1-2u_0}\,,\qquad
\Gamma_K=\frac1{\sqrt{1-3u_0}}\,,\qquad
\frac{\Omega_{\rm(orb)}}{{\Omega_{\rm(ep)}}}=\frac1{\sqrt{1-6u_0}}\,.
\end{eqnarray}

\end{widetext}


\begin{thebibliography}{99}

\bibitem{math37} 
M. Mathisson,
{Acta Phys.\ Polon.} {\bf 6}, 163 (1937).

\bibitem{papa51} 
A. Papapetrou, 
{Proc.\ R.\ Soc.\ A} {\bf 209}, 248 (1951).

\bibitem{tulc59} 
W. Tulczyjew, 
{Acta\ Phys.\ Polon.} {\bf 18}, 393 (1959).

\bibitem{dixon64} 
W.G. Dixon, 
{Nuovo Cimento} {\bf 34}, 317 (1964).

\bibitem{dixon69}
W.G. Dixon, 
{Proc.\ R.\ Soc.\ A} {\bf 314}, 499 (1970).

\bibitem{dixon70}
W.G. Dixon, 
{Proc.\ R.\ Soc.\ A} {\bf 319}, 509 (1970).

\bibitem{dixon73}
W.G. Dixon, 
{Gen.\ Relativ.\ Gravit.} {\bf 4}, 199  (1973).

\bibitem{dixon74}
W.G. Dixon,  
{Phil.\ Trans.\ R.\ Soc.\ A} {\bf 277}, 59 (1974).

\bibitem{dixon_varenna}
W.G. Dixon, in
\textit{Isolated Gravitating System in General Relativity}, 
edited by J. Ehlers (North-Holland, Amsterdam, 1979), p. 156.


\bibitem{ehlers77} 
J. Ehlers  and E. Rudolph,
{Gen.\ Relativ.\ Gravit.} {\bf 8}, 197 (1977).



\bibitem{steinhoff} 
  J.~Steinhoff and D.~Puetzfeld,
Phys.\ Rev.\ D {\bf 86}, 044033 (2012).  


\bibitem{steinhoff2} 
  J.~Steinhoff,   
Ann. Phys. (Berlin) {\bf 523}, 296 (2011).


\bibitem{hinderer}
T. Hinderer, A. Buonanno, A.H. Mrou\'e, D.A. Hemberger, G. Lovelace, H.P. Pfeiffer,  L.E. Kidder, M.A. Scheel, B. Szil\'agyi, N.W. Taylor, and S.A. Teukolsky,
{Phys.\ Rev.\ D} {\bf 88}, 084005 (2013).

\bibitem{bfg}
D. Bini, G. Faye and A. Geralico,
\lq\lq Dynamics of extended bodies in a Kerr spacetime with spin-induced quadrupole tensor,"
in preparation.

\bibitem{porto06}
R.A. Porto,
{Phys.\ Rev.\ D} {\bf 73}, 104031 (2006).


\bibitem{porto08}
R.A. Porto and I.Z. Rothstein, 
{Phys.\ Rev.\ D} {\bf 78}, 044013 (2008).

\bibitem{steinhoff08}
J. Steinhoff, S. Hergt, and G. Sch\"afer, 
{Phys. Rev. D} {\bf 78}, 101503(R) (2008).

\bibitem{hergt08}
S. Hergt and G. Sch\"afer,
{Phys. Rev. D} {\bf 78}, 124004 (2008).

\bibitem{steinhoff09}
J. Steinhoff and G. Sch\"afer, 
{Phys. Rev. D} {\bf 80}, 088501 (2009). 

\bibitem{Levi:2014gsa} 
  M.~Levi and J.~Steinhoff,
  ``Leading order finite size effects with spins for inspiralling compact binaries,''  
arXiv:1410.2601 [gr-qc]. 

\bibitem{Levi:2015msa} 
  M.~Levi and J.~Steinhoff,
  ``An effective field theory for gravitating spinning objects in the post-Newtonian scheme,''  
arXiv:1501.04956 [gr-qc]. 


\bibitem{quadrup_kerr1}
D. Bini and A. Geralico,
Phys.\ Rev.\ D {\bf 89}, 044013 (2014).


\bibitem{quadrup_schw}
D. Bini and A. Geralico, 
{Phys.\ Rev.\ D} {\bf 87}, 024028 (2013).

\bibitem{mashsingh}
B. Mashhoon and D. Singh, 
{Phys.\ Rev.\ D} {\bf 74}, 124006 (2006).

\bibitem{spin_dev_schw} 
D. Bini, A. Geralico, and R.T. Jantzen,
{Gen.\ Relativ.\ Gravit.} {\bf 43}, 959 (2011).


\bibitem{mash1}
B. Mashhoon, 
{Astrophys. Jour.} {\bf 185}, 83 (1973).

\bibitem{mash2}
B. Mashhoon, 
{Astrophys. Jour.} {\bf 216}, 591 (1977).

\bibitem{damour}
T. Damour and J.H. Taylor, 
{Phys.\ Rev.\ D} {\bf 45}, 1840 (1992).

\bibitem{kopeikin}
S.M. Kopeikin, 
Astrophys. Jour. {\bf 467}, L93 (1996).


\bibitem{applegate}
J.H. Applegate and J. Shaham, 
Astrophys. Jour. {\bf 436}, 312 (1994).


\bibitem{doroshenko}
O. Doroshenko, O. L\"ohmer, M. Kramer, A. Jessner, R. Wielebinski, A.G. Lyne, and C. Lange,
Astron. Astrophys. {\bf 379}, 579 (2001).


\bibitem{lazaridis}
K. Lazaridis et al.,
Mon. Not. R. Astron. Soc. {\bf 414}, 3134 (2011).

\bibitem{lanza}
A.F. Lanza and M. Rodon\`o, 
Astron. Astrophys. {\bf 349}, 887 (1999).


\bibitem{MTW}
C.W. Misner, K.S. Thorne, and J.A. Wheeler,
\textit{Gravitation} (Freeman, San Francisco, 1973). 


\bibitem{quadrup_kerr_num}
D. Bini and A. Geralico,
 Classical\ Quantum\ Gravity\ {\bf 31}, 075024 (2014).


\bibitem{dam88}
  T.~Damour and G.~Sch\"afer,
  Nuovo Cimento Soc. Ital. Fis. {\bf 101B}, 127 (1988).


\bibitem{dam-derue1}
T. Damour and N. Deruelle, 
Ann. Inst. Henri Poincar\'e Phys. Theor. {\bf 43}, 107 (1985).


\bibitem{dam-derue2}
T. Damour and N. Deruelle,  
Ann. Inst. Henri Poincar\'e Phys. Theor. {\bf 44}, 263 (1986).


\bibitem{wex}
G.~Sch\"afer and N.~Wex,
Phys. Lett. A {\bf 174}, 196 (1993). 
 



\end{thebibliography}
\end{document}